%% file: compound_ifc7.tex
\documentclass[12pt]{article}
\usepackage{latexsym,graphics,graphicx,amssymb,amsmath, color}
\makeatletter
\title{}
\author{}
\date{}
\newcommand{\be}{\begin{equation}}
\newcommand{\ee}{\end{equation}}

\newcommand{\nn}{\nonumber}
\newcommand{\bc}{\begin{center}}
\newcommand{\ec}{\end{center}}
\newcommand{\bfl}{\begin{flushleft}}
\newcommand{\efl}{\end{flushleft}}

\newcommand{\beqa}{\begin{eqnarray}}
\newcommand{\eeqa}{\end{eqnarray}}
\newcommand{\beqan}{\begin{eqnarray*}}
\newcommand{\eeqan}{\end{eqnarray*}}
\newcommand{\beq}{\begin{equation}}
\newcommand{\eeq}{\end{equation}}

\newcommand{\lbr}{\left \{ }
\newcommand{\rbr}{\right \} }
\newcommand{\Lbr}{\left [}
\newcommand{\Rbr}{\right ]}
\newcommand{\lp}{\left (}
\newcommand{\rp}{\right )}
\newcommand{\df}{\triangleq}
\newcommand{\real}{{\mathbb{R}}}

\newcommand{\E}{{\mathbb{E}}}

\newtheorem{prop}{Proposition}

\newtheorem{claim}{Claim}

\newcommand{\ga}{\gamma}
\newcommand{\de}{\delta}
\newcommand{\om}{\omega}
\newcommand{\La}{\Lambda}
\newcommand{\la}{\lambda}
\newcommand{\A}{\mathcal{A}}

\newtheorem{thm}{Theorem}
\newtheorem{corollary}[thm]{Corollary}
\newtheorem{lem}[thm]{Lemma}

\author{Adnan Raja, Vinod  M.\ Prabhakaran and Pramod Viswanath}
\title{The Two User Gaussian Compound Interference Channel}
\date{\today}

\begin{document}
\maketitle

\begin{abstract}
We introduce the two user  finite state compound Gaussian
interference channel and characterize its capacity region to within
one bit. The main contributions involve both novel inner and outer
bounds. The inner bound is multilevel superposition coding but
the decoding of the levels is opportunistic, depending on the
channel state. The genie aided outer bound is motivated by the
typical error events of the achievable scheme.
\end{abstract}

\section{Introduction}
\label{sec:intro}

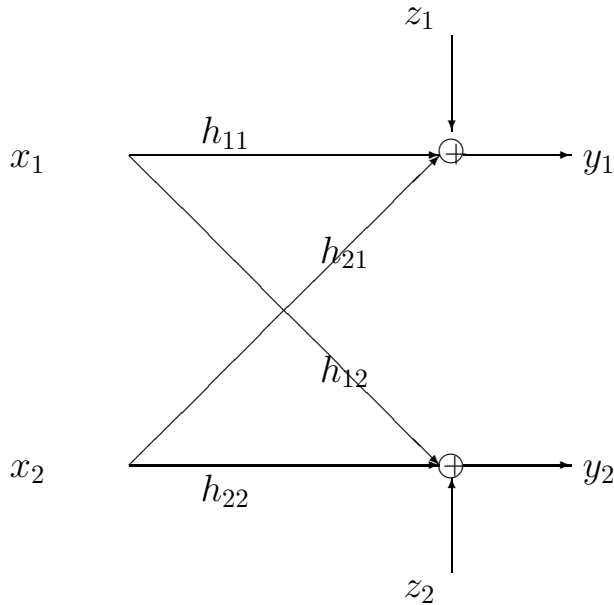
\begin{figure}[h]
\begin{center}
\scalebox{1.0}{\input{GIC.latex}}
\end{center}
\caption{The two user Gaussian interference channel. }
\label{fig-GICmodel}
\end{figure}

The focus of this paper is the communication scenario depicted in
Figure~\ref{fig-GICmodel}. Two transmitter-receiver pairs
communicate reliably in the face of interference. The discrete time
complex baseband model is: \beqa y_1[m] & = &
h_{11}x_1[m] + h_{21}x_2[m] + z_1[m],\\
y_2[m]&=&h_{12}x_1[m]+h_{22}x_2[m]+z_2[m].  \eeqa Here $m$ is the
time index, $y_k$ is the signal at receiver $k$ while $x_k$ is the
signal sent out by the transmitter $k$ (with $k=1,2$). The noise
sequences $\lbr z_1[m], z_2[m]\rbr_m$ are memoryless complex
Gaussian with zero mean and unit variance. The transmitters are
subject to average power constraints: \beq \sum_{m=1}^N |x_k[m]|^2
\leq NP_k, \quad k=1,2, \quad \forall N \geq 1. \eeq The complex
parameters $\lbr h_{k\ell }, \ell =1,2, k=1,2\rbr$ model the channel
coefficients between the pairs of transmitters and receivers. They
do not vary with time but the transmitters and receivers have
different information about them:
\begin{itemize}
\item Receiver $k$ is exactly aware of the two channel coefficients $h_{1k}, h_{2k}$;
this models {\em coherent} communication.
\item Transmitters are only {\em coarsely} aware of the channel coefficients: the
transmitters know that the channel coefficients belong to a finite
set. Specifically, both the transmitters know that \beq
(h_{1k},h_{2k}) \in \A_{k}, \quad k=1,2. \label{eq:sets} \eeq This
models potential partial feedback to the transmitters regarding the
channel coefficients.
\end{itemize}
A more general compound channel model allows for all four channel
parameters to jointly take on different choices:
\beq\label{eq:sets_general} (h_{11},h_{12},h_{21}, h_{22}) \in \A.
\eeq However, since the receivers do not cooperate in the
interference channel, it turns out that the setting in
Equation~\eqref{eq:sets_general} is no more general than the one in
Equation~\eqref{eq:sets}. This is explored in
Section~\ref{sec:Gaussian}.

The key problem of interest is the characterization of the capacity
region: the set of rate pairs at which arbitrarily reliable
communication between the two transmitter-receiver pairs. The
``compound" aspect of the channel is in insisting that the receivers
be able to decode the messages of interest with arbitrarily high
probability, no matter which of the finite states the channel
coefficients take on. Our main result is a characterization of the
capacity region up to one bit.

A special instance of the problem studied here is the classical two
user Gaussian interference channel: in a recent work, Etkin, Tse and
Wang \cite{ETW07} showed that a {\em single} superposition coding
scheme (a specific choice among the broad class of schemes first
identified by Han and Kobayashi \cite{HK81}) achieves performance
within one bit of the capacity region. The transmission involved
splitting the data into two parts --  one {\em public} and the other
{\em private} --
 and linearly superposing them.
The idea is that the public data stream is decoded by both the
receivers while the private data stream only by the receiver of
interest. The  key identity of the proposed superposition scheme is
the following: the power allocated to the private stream is such
that it appears at exactly the same level as the background noise at
the unintended receiver (the idea is that since the private data
stream is being treated as noise at the unintended receiver, there
is no extra incentive to reduce its level even further than that of
the additive noise). A novel outer bound developed in \cite{ETW07}
showed that this simple superposition scheme is within one bit of
the capacity region.

Implementation of the specific superposition scheme proposed above
requires each  transmitter to be aware of the interference level it
is  causing to the unintended receiver. In the context of the
compound channel being studied here, the transmitter is not aware of
the interference level; this poses an obstacle to adopting the idea
of appropriately choosing the power level of the private data
stream. One possibility could be to set the power level of the
private data stream based on the strongest interfering link level
(among the set of possible choices) -- this would ensure that it is
only received {\em below} noise level when the interfering link
level takes on the other possible choices. However, this approach
might be too pessimistic and its closeness to optimality is unclear.

We circumvent this problem by proposing the following novel twist to
the general superposition coding scheme. Our main idea is best
described when the interference links ($h_{12}$ and $h_{21}$ in
Figure~\ref{fig-GICmodel}) take on only two possible values and the
direct links are fixed  (i.e., the sets $\A_{1}$ and $\A_{2}$ have
cardinality of two, cf.\ Equation~\eqref{eq:sets}). We now superpose
{\em three} data streams at each transmitter. Two of them, public
and private, are as earlier: all receivers  in all channel states
decode the public message while only the receiver of interest
decodes the private message (no matter the channel state, again).
The novelty is in the third data stream that we will call {\em
semi-public}: this data stream is decoded by the unintended receiver
only when the interference link is the stronger of the two choices
(and treated as noise otherwise). As such, this data stream is
neither fully private nor public (the unintended receiver either
treats it as noise or decodes it based on the channel state) and the
nomenclature is chosen to highlight this feature.

The power split rule is the following: the power of the private
stream is set such that at the higher of the interference link
levels, it is received at the unintended receiver at the same level
as the additive noise. The power of the semi-public data stream is
set such that  it is received at the unintended receiver at the same
level as the additive noise only when the interference link level is
at the lower of the two possible choices. The rationale is that the
semi-public data stream is not  decoded only when the interference
link level is at the lower of the two possible choices, and thus it
can transmit higher  power than if its power is restricted by the
higher of the interference link levels. This approach scales
naturally when the interference link levels can take on more than
two possible choices (the number of splits of the data stream is one
more than the cardinality of the set of possible choices).

We derive novel outer bounds to show that our simple achievable
scheme is within one bit of the capacity region. Our outer bounds
are genie aided and are based on the clues provided by the {\em
typical error events} in the achievable scheme.  This approach sheds
operational insight into the nature of the outer bounds even in the
noncompound version (thus eliminating the ``guesswork" involved in
the derivation, cf.\ Section IV of \cite{TT07}).

The paper is organized as follows: we start with a simple two-state
compound interference channel. In this setting, both the direct and
interference link levels can take on only one of two possible values
(so the sets $\A_{1}$ and $\A_{2}$ have cardinality two). Using a
somewhat abstract setting (described in Section~\ref{sec:model})
that features the Gaussian problem of interest as a special case, we
present our main results (both inner and outer bounds) for this
two-state compound interference channel. Our definition of the
abstract setting is motivated by that chosen in \cite{TT07} and
could be viewed as a natural compound version of the interference
channel studied by Telatar and Tse \cite{TT07}. This is done in
Section~\ref{sec:result}. We discuss the insights garnered from
these results in the context of the simpler noncompound interference
channel in Section~\ref{sec:discuss}. Next, we are ready to set up
the model and describe the solution the more general finite state
interference channel; we do this first in the abstract setting
(Section~\ref{sec:gen}) followed by specializing to the Gaussian
scenario of interest (Section~\ref{sec:Gaussian}).

\section{Model}
\label{sec:model}

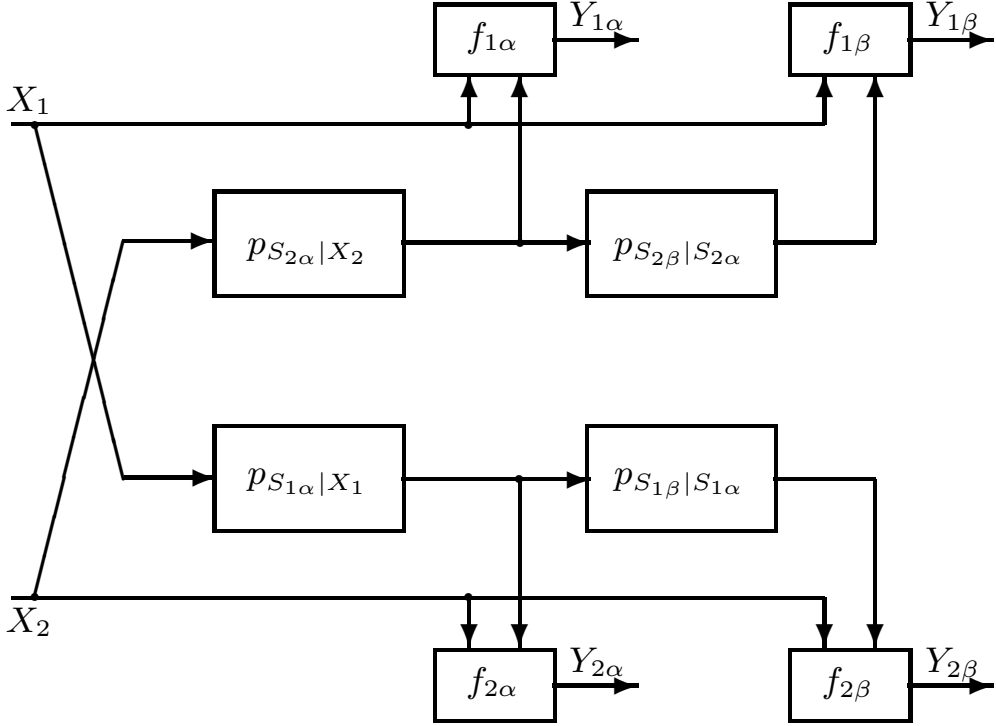
\begin{figure}[h]
\begin{center}
\scalebox{1.5}{\input{CompoundIC.latex}}
\end{center}
\caption{A two-state compound channel model.} \label{fig-cICmodel}
\end{figure}

Consider the two-user, two-state compound memoryless interference
channel depicted in Figure~\ref{fig-cICmodel}.  There are two
transmitters
 which want to reliably communicate independent messages to two
corresponding receivers. The input to the channel from the first
transmitter at any discrete time $X_1\in{\mathcal X}_1$ passes
through a degraded discrete memoryless broadcast channel: the two
outputs of the degraded broadcast  channel are
$S_{1\alpha}\in{\mathcal S}_1$ and (the degraded version)
$S_{1\beta}\in{\mathcal S}_1$.  Similarly, at any time, the input to
the channel from transmitter 2 $X_2\in{\mathcal X}_2$ produces
$S_{2\alpha}\in{\mathcal S}_2$ and a degraded version
$S_{2\beta}\in{\mathcal S}_2$ of it.
The channel to any one of the two receivers is decided by the {\em
state} of that receiver: here there are only two states $\alpha$ and
$\beta$. Once the state is decided, it is fixed for the entire
duration of communication.
 When
the first receiver  is in state $\alpha$, the output at any time is
\beq Y_{1\alpha}=f_{1\alpha}(X_1,S_{2\alpha}) \in{\mathcal Y}_1.
\eeq Similarly, when the first receiver is in state $\beta$, the
output at any time is \beq Y_{1\beta}=
f_{1\beta}(X_1,S_{2\beta})\in{\mathcal Y}_1. \eeq Here $f_{1\alpha}$
and $f_{1\beta}$ are deterministic functions such that for every
$x_1\in{\mathcal X}_1,s_2\in{\mathcal S}_2$, and
$\eta=\alpha,\beta$, the following function is invertible:
\begin{align*}
f_{1\eta}(x_1,.):{\mathcal S}_2\rightarrow {\mathcal Y}_1.
\end{align*}

Likewise, the outputs of user~2 under the two possible states the
channel to it can take are defined using similar deterministic
functions $f_{2\alpha}$ and $f_{2\beta}$.

We allow each receiver to be in potentially different states, and
they are both aware of the state they are in. A pair of
communication rates $(R_1,R_2)$ is said to be {\em achievable} if
for every $\epsilon>0$, there are block length $n$ encoders, \beq
\text{enc}_k:\{1,\ldots,M_k\}\rightarrow{\mathcal X}_k^n, M_k\geq
2^{n(R_k-\epsilon)},\quad k=1,2, \eeq and decoders
\beq\text{dec}_{k\eta}:{\mathcal
Y}_k^n\rightarrow\{1,\ldots,M_k\},\quad k=1,2,\quad \eta=\alpha,
\beta, \eeq such that \beq\frac{1}{M_1M_2}\sum_{m_1,m_2}
\text{Pr}\lp \text{dec}_{k,\eta}(Y_{k\eta}^n) = m_k,\; k=1,2,\;
\eta=\alpha,\beta | X_k^n=\text{enc}_k(m_k),\; k=1,2\rp \geq
1-\epsilon. \eeq We are interested in the capacity region ${\mathcal
C}$, which is the set of all achievable $(R_1,R_2)$ pairs. We can
make a few observations:
\begin{itemize}
\item The channel described here can be thought of as a natural generalization
of that studied in \cite{TT07}.
\item An important special case occurs when the
channels $p_{S_{k\alpha}|X_k}$ and $p_{S_{k\beta}|S_{k\alpha}}$ are
{\em deterministic} for both $k=1,2$. This channel is a compound
version of the deterministic channel considered by El Gamal and
Costa~\cite{EC82} with the interference in state~$\beta$ being a
deterministic function of the interference in state~$\alpha$.
\item The compound Gaussian interference channel, with the cardinality of both the sets $\A_1$ and $\A_2$ restricted to $2$ (in the notation introduced in Section~\ref{sec:intro}), is a special instance of the model
in Figure~\ref{fig-cICmodel}. We start with a compound Gaussian
interference channel with
\begin{align}
(h_{11},h_{21})&\in \lbr (h_{11\alpha},h_{21\alpha}), (h_{11\beta},h_{21\beta}) \rbr, \nn\\
(h_{22},h_{12})&\in \lbr (h_{22\alpha},h_{12\alpha}),
(h_{22\beta},h_{12\beta}) \rbr. \nn
\end{align}
Further, without loss of generality, we can assume that \beqa
|h_{21\alpha}| &\geq& |h_{21\beta}|, \\
|h_{12\alpha}| &\geq& |h_{12\beta}|. \eeqa With the following
assignment, we see that the model in Figure~\ref{fig-cICmodel} can
capture the Gaussian model in Figure~\ref{fig-GICmodel}:
\begin{align}
S_{1\alpha} &= h_{12\alpha}X_{1}+Z_{2}, \\
S_{1\beta} &=
\frac{h_{12\beta}}{h_{12\alpha}}S_{1\alpha}+\lp 1 - \left| \frac{h_{12\beta}}{h_{12\alpha}}\right|^{2} \rp^{1/2} Z_{2}^{\prime}, \\
S_{2\alpha} &= h_{21\alpha}X_{2}+Z_{1}, \\
S_{2\beta} &= \frac{h_{21\beta}}{h_{21\alpha}}S_{2\alpha}+\lp 1 -
\left|\frac{h_{21\beta}}{h_{21\alpha}}\right|^{2} \rp^{1/2}
Z_{1}^{\prime},
\end{align}
\begin{align}
Y_{1\alpha} &=f_{1\alpha}(X_{1},S_{2\alpha})= h_{11\alpha}X_{1}+S_{2\alpha}, \\
Y_{1\beta} &=f_{1\beta}(X_{1},S_{2\beta})= h_{11\beta}X_{1}+S_{2\beta}, \\
Y_{2\alpha} &=f_{2\alpha}(X_{2},S_{1\alpha})= h_{22\alpha}X_{2}+S_{1\alpha}, \\
Y_{2\beta} &=f_{2\beta}(X_{2},S_{1\beta})=
h_{22\beta}X_{2}+S_{1\beta}.
\end{align}
Here $Z_{1}, Z_{1}^{\prime}, Z_{2}$ and $Z_{2}^{\prime}$ are
independent complex Gaussian  random variables with unit variance.
\end{itemize}

\section{Main Result}\label{sec:result}

Our main results on the $2$-state compound interference channel  are
the following.
\begin{itemize}
\item We first show the performance of an achievable scheme and hence characterize an inner-bound;
\item next, we give an outer-bound to the capacity region and
quantify the gap between the outer-bound and the achievable scheme;
\item specializing to the compound deterministic  interference channel, we
completely characterize the capacity region;
\item specializing to the compound Gaussian  interference channel, we characterize
the capacity region up to a gap of 1 bit (at all operating SNR
values and all channel parameter values).
\end{itemize}

\subsection{Inner-bound: Achievable Scheme}

The achievable scheme is characterized by ${\mathcal P}$, the set
of random variables
\beq
({Q},{X}_1,U_{1\alpha},U_{1\beta},{X}_2,U_{2\alpha},U_{2\beta}),
\eeq
such that  the following Markov chain is satisfied:
\begin{align*}
U_{1\beta}-U_{1\alpha}-X_1-Q-X_{2}-U_{2\alpha}-U_{2\beta}.
\end{align*}
Alternatively,  the joint probability distribution function factors as
\begin{align}
p(q,x_{1,},u_{1\alpha},u_{1\beta},x_{2},u_{2\alpha},u_{2\beta}) =&
p(q)p(x_{1}|q)p(u_{1\alpha}|x_1)p(u_{1\beta}|u_{1\alpha})
p(x_{2}|q)p(u_{2\alpha}|x_2)p(u_{2\beta}|u_{2\alpha}).
\end{align}

Our achievable scheme is a multilevel superposition coding one
and can be viewed as a generalization of the two-level
superposition coding scheme of Chong et al.\ \cite{CMG06}. The
random coding method can be intuitively described as follows, using
the ``cloud-center" analogy from Cover and Thomas (Section~14.6.3, \cite{CT91});
a formal statement and its proof follow later.  The random variables
$U_{1\beta}$ and $U_{2\beta}$ are used to generate the {\em
outermost} code books (with rate $R_{1\beta}$ and $R_{2\beta}$,
respectively) for the two users. These messages encoded via these
code books are decoded  by both receivers and, as such, can be
interpreted as {\em public} information. Next, the random variables
$U_{1\alpha}$ and $U_{2\alpha}$ are  used to generate  the next
level of code books (with rate $R_{1\alpha}$ and $R_{2\alpha}$,
respectively). The messages encoded via these code books
 are decoded by the receiver with stronger interference
 (i.e.~$\text{Rx}_{k\alpha}$) but treated as noise by the receiver
 with weaker interference (i.e.~$\text{Rx}_{k\beta}$); as such, these
 messages can be viewed as {\em semi-public} information.
Finally, the messages encoded via the inner most code books (rates $R_{1p}$
and $R_{2p}$) are only decoded by the receiver of interest; thus this
constitutes {\em private } information.

Given $P\in{\mathcal P}$, we define the six-dimensional region
\beq
{\mathcal R}_{in}^{(6)}(P)
\df \lbr
\lp R_{1p},R_{1\alpha},R_{1\beta},R_{2p},R_{2\alpha},R_{2\beta}\rp :
\text{satisfying}~\eqref{ibeq1b1}-\eqref{ibeqnn2}\label{eq:Rin6}
\rbr. \eeq {\footnotesize
\begin{align}
R_{1p}&\leq
I(Y_{1\beta};X_{1}|U_{1\alpha},U_{2\beta},Q)&=
\ga_{11} \label{ibeq1b1}\\
R_{2\beta}+R_{1p}&\leq
I(Y_{1\beta};X_{1},U_{2\beta}|U_{1\alpha},Q)&=
\ga_{12} \label{ibeq1b2} \\
R_{1\alpha}+R_{1p}&\leq
I(Y_{1\beta};X_{1}|U_{1\beta},U_{2\beta},Q)&=
\ga_{13}  \\
R_{2\beta}+R_{1\alpha}+R_{1p}&\leq
I(Y_{1\beta};X_{1},U_{2\beta}|U_{1\beta},Q)&=
\ga_{14}  \\
R_{1\beta}+R_{1\alpha}+R_{1p} &\leq
I(Y_{1\beta};X_{1}|U_{2\beta},Q)&=
\ga_{15}  \\
R_{2\beta}+R_{1\beta}+R_{1\alpha}+R_{1p} &\leq
I(Y_{1\beta};X_{1},U_{2\beta}|Q)&=
\ga_{16}  \label{ibeq1b6}
\end{align}

\begin{align}
R_{1p} &\leq
I(Y_{1\alpha};X_{1}|U_{1\alpha},U_{2\alpha},Q) &= \de_{11} \label{ibeq1a1}\\
R_{2\alpha}+R_{1p} &\leq
I(Y_{1\alpha};X_{1},U_{2\alpha}|U_{1\alpha},U_{2\beta},Q) &= \de_{12} \\
R_{2\beta}+R_{2\alpha}+R_{1p} &\leq
I(Y_{1\alpha};X_{1},U_{2\alpha}|U_{1\alpha},Q) &= \de_{13} \label{ibeq1a3}  \\
R_{1\alpha}+R_{1p} &\leq
I(Y_{1\alpha};X_{1}|U_{1\beta},U_{2\alpha},Q) &= \de_{14}  \\
R_{2\alpha}+R_{1\alpha}+R_{1p} &\leq
I(Y_{1\alpha};X_{1},U_{2\alpha}|U_{1\beta},U_{2\beta},Q) &= \de_{15}  \\
R_{2\beta}+R_{2\alpha}+R_{1\alpha}+R_{1p} &\leq
I(Y_{1\alpha};X_{1},U_{2\alpha}|U_{1\beta},Q) &= \de_{16}  \\
R_{1\beta}+R_{1\alpha}+R_{1p} &\leq
I(Y_{1\alpha};X_{1}|U_{2\alpha},Q) &= \de_{17}  \\
R_{2\alpha}+R_{1\beta}+R_{1\alpha}+R_{1p} &\leq
I(Y_{1\alpha};X_{1},U_{2\alpha}|U_{2\beta},Q) &= \de_{18}   \\
R_{2\beta}+R_{2\alpha}+R_{1\beta}+R_{1\alpha}+R_{1p} &\leq
I(Y_{1\alpha};X_{1},U_{2\alpha}|Q) &= \de_{19}  \label{ibeq1a9}
\end{align}

\begin{align}
R_{2p}&\leq
I(Y_{2\beta};X_{2}|U_{2\alpha},U_{1\beta},Q) &= \ga_{21} \label{ibeq2b1}\\
R_{1\beta}+R_{2p} &\leq
I(Y_{2\beta};X_{2},U_{1\beta}|U_{2\alpha},Q) &= \ga_{22} \label{ibeq2b2} \\
R_{2\alpha}+R_{2p} &\leq
I(Y_{2\beta};X_{2}|U_{2\beta},U_{1\beta},Q) &= \ga_{23}  \\
R_{1\beta}+R_{2\alpha}+R_{2p} &\leq
I(Y_{2\beta};X_{2},U_{1\beta}|U_{2\beta},Q) &= \ga_{24}  \\
R_{2\beta}+R_{2\alpha}+R_{2p} &\leq
I(Y_{2\beta};X_{2}|U_{1\beta},Q) &= \ga_{25}  \\
R_{1\beta}+R_{2\beta}+R_{2\alpha}+R_{2p} &\leq
I(Y_{2\beta};X_{2},U_{1\beta}|Q) &= \ga_{26}  \label{ibeq2b6}
\end{align}

\begin{align}
R_{2p} &\leq
I(Y_{2\alpha};X_{2}|U_{2\alpha},U_{1\alpha},Q) &= \de_{21} \label{ibeq2a1}\\
R_{1\alpha}+R_{2p} &\leq
I(Y_{2\alpha};X_{2},U_{1\alpha}|U_{2\alpha},U_{1\beta},Q) &= \de_{22} \\
R_{1\beta}+R_{1\alpha}+R_{2p} &\leq
I(Y_{2\alpha};X_{2},U_{1\alpha}|U_{2\alpha},Q) &= \de_{23} \label{ibeq2a3} \\
R_{2\alpha}+R_{2p} &\leq
I(Y_{2\alpha};X_{2}|U_{2\beta},U_{1\alpha},Q) &= \de_{24}  \\
R_{1\alpha}+R_{2\alpha}+R_{2p} &\leq
I(Y_{2\alpha};X_{2},U_{1\alpha}|U_{2\beta},U_{1\beta},Q) &= \de_{25}  \\
R_{1\beta}+R_{1\alpha}+R_{2\alpha}+R_{2p} &\leq
I(Y_{2\alpha};X_{2},U_{1\alpha}|U_{2\beta},Q) &= \de_{26}  \\
R_{2\beta}+R_{2\alpha}+R_{2p} &\leq
I(Y_{2\alpha};X_{2}|U_{1\alpha},Q) &= \de_{27}  \\
R_{1\alpha}+R_{2\beta}+R_{2\alpha}+R_{2p} &\leq
I(Y_{2\alpha};X_{2},U_{1\alpha}|U_{1\beta},Q) &= \de_{28}   \\
R_{1\beta}+R_{1\alpha}+R_{2\beta}+R_{2\alpha}+R_{2p} &\leq
I(Y_{2\alpha};X_{2},U_{1\alpha}|Q) &= \de_{29}  \label{ibeq2a9}
\end{align}

\begin{eqnarray}
R_{1p}+R_{1\alpha}+R_{1\beta} &\geq& 0 \label{ibeqnn1}\\
R_{2p}+R_{2\alpha}+R_{2\beta} &\geq& 0. \label{ibeqnn2}
\end{eqnarray}}

We define the two-dimensional region,

\begin{align}
{\mathcal R}_{in}(P) \df &\{ (R_{1},R_{2}): \nn \\
&\;\;R_1=R_{1p}+R_{1\alpha}+R_{1\beta}, \nonumber\\
&\;\;R_2=R_{2p}+R_{2\alpha}+R_{2\beta}, \nonumber \\
&\;\;(R_{1p},R_{1\alpha},R_{1\beta},R_{2p},R_{2\alpha},R_{2\beta})
\in {\mathcal R}_{in}^{(6)}(P) \}.
\end{align}

In other words ${\mathcal R}_{in}(P)$ is the projection of the
six-dimensional polytope ${\mathcal R}^{(6)}_{in}(P)$. One approach
to take the projection, is to do the Fourier-Motzkin
elimination, as done for the basic superposition coding scheme in
the context of the regular (noncompound) interference channel~\cite{CMG06}.
 Doing this explicitly is rather cumbersome as the inequalities here are much more in number than the inequalities that were handled by Chong et al.\ in~\cite{CMG06}.

\begin{thm} \label{thm:Main1}
 The capacity region ${\mathcal C}$ satisfies
 \beq {\mathcal C} \supseteq \bigcup_{P\in{\mathcal P}} {\mathcal
R}_{in}\lp P \rp.\eeq
\end{thm}

\noindent\emph{Proof:} A formal description of the achievable scheme
and the proof of this theorem are available in Section~\ref{sec:thm1}.
\hfill$\Box$

Particularizing, we restrict ourselves to a subset of
${\mathcal P}$ defined as follows. Given random variables
$(Q,X_1,X_2)$ such that $X_1$ and $X_2$ are conditionally
independent when conditioned on $Q$, we define random variables
$U_{1\alpha}$ and $U_{1\beta}$ which take values in ${\mathcal
S}_1$, and $U_{2\alpha}$ and $U_{2\beta}$ which take values in
${\mathcal S}_2$. They are jointly distributed with $(Q,X_1,X_2)$
according to the conditional distribution
\begin{align}
p(u_{1\alpha},u_{1\beta},u_{2\alpha},u_{2\beta}|q,x_1,x_2)  &=& \nn\\
&p_{S_{1\alpha}|X_1}(u_{1\alpha}|x_1)p_{S_{1\beta}|S_{1\alpha}}(u_{1\beta}|u_{1\alpha})
p_{S_{2\alpha}|X_2}(u_{2\alpha}|x_2)p_{S_{2\beta}|S_{2\alpha}}(u_{2\beta}|u_{2\alpha}).\label{eq:aux}
\end{align}
Note that, conditioned on $Q$, we have the following two Markov
chains, with the sets of random variables involved in the two chains
being conditionally independent.
\begin{align*}U_{1\beta}-U_{1\alpha}-X_1-S_{1\alpha}-S_{1\beta}\\
U_{2\beta}-U_{2\alpha}-X_2-S_{2\alpha}-S_{2\beta}.
\end{align*}
Our choice is motivated by the choice in the paper by Telatar and
Tse~\cite{TT07}. Every member of this family is uniquely
determined by joint random variables $\lp Q,X_{1},X_{2} \rp$ such
that $X_{1}-Q-X_{2}$ is a Markov chain. We will henceforth denote
the corresponding regions $R_{in}^{(6)} \lp P \rp$ by
$R_{in}^{(6)}\lp Q,X_{1},X_{2} \rp$ and $R_{in} \lp P \rp$ by
$R_{in}\lp Q,X_{1},X_{2} \rp$.  We now have the natural result:

\begin{corollary}
\beq {\mathcal C} \supseteq \bigcup_{Q,X_1,X_2} {\mathcal R}_{in}\lp
Q,X_1,X_2 \rp,
\label{eq:maincorollary}
\eeq
 where the union is over all $(Q,X_1,X_2)$ such
that $X_1 - Q - X_2$ is a Markov chain.
\end{corollary}

\noindent\emph{Proof:} Follows directly from Theorem~\ref{thm:Main1}.
\hfill$\Box$

We observe that the Fourier-Motzkin elimination procedure to implement
the projection operation in obtaining
$\mathcal R_{in}({Q},{X}_{1},{X}_{2})$ would yield only a finite
set of inequalities. Further, the right hand sides of these
inequalities would be linear functions of $p(q)$ and for a fixed
$Q=q_0$ the right hand sides form a closed set of finite dimensions.
Thus, by Carath\`{e}odory's theorem, we can conclude that the
cardinality of $Q$ can taken to be finite without loss of generality
in the union in Equation~\eqref{eq:maincorollary}.

\subsection{Outer-bound}

\begin{thm}\label{thm:Main2}
 For every $({Q},{X}_1,{X}_2)$ such
that ${X}_1 - {Q} - {X}_2$ is a Markov chain, there is a region
${\mathcal R}_{out}({Q},{X}_1,{X}_2)\subseteq {\mathbb R}_+^2$ such
that the following are true:
\begin{enumerate}
\item[(i)]
\beq {\mathcal C} \subseteq
                  \bigcup_{{Q},{X}_1,{X}_2} {\mathcal R}_{out}({Q},{X}_1,{X}_2),\eeq
where the union is over all $({Q},{X}_1,{X}_2)$  such that ${X}_1 -
{Q} - {X}_2$ is a Markov chain.
\item[(ii)]
If $(R_1,R_2)\in{\mathcal R}_{out}(Q,X_1,X_2)$, then
$(R_1-\Delta_1,R_2-\Delta_2)\in{\mathcal R}_{in}(Q,X_1,X_2)$, where
\begin{align}
\Delta_1(Q,X_1,X_2)&=\max(I(X_2;S_{2\alpha}|U_{2\alpha}),
   I(X_2;S_{2\beta}|U_{2\beta})), \\
\Delta_2(Q,X_1,X_2)&=\max(I(X_1;S_{1\alpha}|U_{1\alpha}),
   I(X_1;S_{1\beta}|U_{1\beta})),
\end{align}
in which the random variables are jointly distributed according to
\eqref{eq:aux} and the channel conditional distributions.
\end{enumerate}
\end{thm}
\noindent\emph{Proof:} Part(i) is proved in
Section~\ref{sec:thm2-1}. Part(ii) is proved in
Section~\ref{sec:thm2-2}. \hfill$\Box$

The set ${\mathcal R}_{out}(Q,X_1,X_2)$ is defined in
Section~\ref{sec:thm2-1}. Our definition is motivated by the
\emph{external representation} of ${\mathcal R}_{in}(Q,X_1,X_2)$
that we obtain in Section~\ref{sec:extRep}.

\subsection{Special Cases}

Our model captures two important special cases:
\begin{itemize}
\item the compound deterministic
interference channel;
\item the compound Gaussian interference channel,
\end{itemize}
as discussed in Section \ref{sec:model}. Thus our results apply
to these cases (readily for the deterministic channel, and with an
appropriate approximation result to the continuous alphabet Gaussian channel).
 Moreover, the structure afforded by these special
cases allows us to derive further insight into the nature of the
general results derived earlier.

\subsubsection{Compound Deterministic Interference Channel}

In this instance, the capacity region is exactly  described.
\begin{corollary} For the deterministic compound interference channel, the
inner bound in Theorem~\ref{thm:Main1} is the capacity region.
\end{corollary}

\noindent{\em Proof}: The proof is elementary. When the channel is
deterministic, we see that the  gap claimed by
Theorem~\ref{thm:Main2} \beq
\Delta_1(Q,X_1,X_2)=\Delta_2(Q,X_1,X_2)=0. \eeq This completes the
proof. \hfill $\Box$

\subsubsection{$2$-state Compound Gaussian Interference Channel}

For the Gaussian version, we can characterize the capacity to within
one-bit.
\begin{corollary} For the $2$-state compound Gaussian interference channel, the
achievable region of Theorem~\ref{thm:Main1} is within at most one
bit of the capacity region.
\end{corollary}

\noindent{\em Proof}: For the Gaussian channel, each of the mutual
information terms in the expressions for $\Delta_1(Q,X_1,X_2)$ and
$\Delta_2(Q,X_1,X_2)$ can be upper bounded by 1 bit. To see this,
note that $S_{1\alpha}=h_{1\alpha}X_1+N_{1\alpha}$ and
$U_{1\alpha}=h_{1\alpha}X_1+N_{1\alpha}^\prime$, where $N_{1\alpha}$
and $N_{1\alpha}^\prime$ are independent and identically distributed
memoryless complex Gaussian random variables. Hence
\begin{align*}
I(X_1;S_{1\alpha}|U_{1\alpha}) &= h(S_{1\alpha}|U_{1\alpha}) -
h(N_{1\alpha})\\ &\leq h(S_{1\alpha}-U_{1\alpha}) - h(N_{1\alpha}) =
1.
\end{align*}
Similarly, \beqa
I(X_1;S_{1\beta}|U_{1\beta})&\leq& 1, \\
I(X_2;S_{2\alpha}|U_{2\alpha})&\leq& 1,\\
I(X_2;S_{2\beta}|U_{2\beta})&\leq& 1. \eeqa \hfill $\Box$

Additionally, we can use Gaussian code books to get to within one
bit of the capacity.
\begin{corollary}\label{cor:gcodebook} For the $2$-state compound Gaussian interference channel,
\beq {\mathcal C} \subseteq
                   {\mathcal R}_{out}({Q}^{*},{X}_1^{*},{X}_2^{*}),\eeq
where
$Q^{*}=1,\;X_{1}^{*}\sim\mathcal{CN}(0,P_{1}),\;X_{2}^{*}\sim\mathcal{CN}(0,P_{2})$.

This implies that ${\mathcal R}_{in}({Q}^{*},{X}_1^{*},{X}_2^{*})$
is within one-bit of the capacity region ${\mathcal C}$ of the
2-state Gaussian compound interference channel.
\end{corollary}

\noindent{\em Proof}: See Section \ref{sec:proofgcodebook}. \hfill$\Box$

\section{An Achievable Scheme}
\label{sec:ib}

We will present a natural, and novel, achievable scheme first. We
will evaluate the set of reliable communication rates using this
strategy and hence characterize an inner bound to the capacity
region; this will complete the proof of Theorem~\ref{thm:Main1}.
Next, we will see some important geometric properties of the
achievable rate region.

\subsection{Proof Of Theorem~\ref{thm:Main1}} \label{sec:thm1}
Our coding scheme is a natural generalization of the scheme of Chong
et  al.\ \cite{CMG06}.  Since there are two possible states for both
receivers, each encoder now sends two sets of common information, with
the receivers opportunistically decoding the common information (depending
on the state). we choose the random
variables corresponding to the two sets of common information in a
degraded manner, following the same ordering of degradedness of
the
interferences under the two states (c.f.\  \beq
X_k-S_{k\alpha}-S_{k\beta},\quad k=1,2). \eeq

Fix a $P\in{\mathcal P}$.

\subsubsection*{Codebook Generation}

Generate a codeword $Q^{n}$ of length $n$, generating each element
i.i.d.\ according to $\Pi_{i=1}^{n}p(q_{i})$. For the codeword
$Q^{n}$, generate $2^{nR_{1\beta}}$ independent codewords
\beq U_{1\beta}^{n}(j_{1}),\quad
j_{1}\in\{1,2,\cdots,2^{nR_{1\beta}}\},\eeq generating each element
i.i.d.\ according to $\Pi_{i=1}^{n}p(u_{1\beta i}|q_{i})$. For each
of the codewords $U_{1\beta}^{n}(j_{1})$, generate $2^{nR_{1\alpha}}$
independent codewords \beq U_{1\alpha}^{n}(j_{1},k_{1}),\quad
k_{1}\in\{1,2,\cdots,2^{nR_{1\alpha}}\},\eeq generating each element
i.i.d.\ according to $\Pi_{i=1}^{n}p(u_{1\alpha i}|u_{1\beta
i},q_{i})$. For each of the codewords $U_{1\alpha}^{n}(j_{1},k_{1})$,
generate $2^{nR_{1p}}$ independent codewords
\beq X_{1}^{n}(j_{1},k_{1},l_{1}),\quad
l_{1}\in\{1,2,\cdots,2^{nR_{1p}}\},\eeq generating each element
i.i.d.\ according to $\Pi_{i=1}^{n}p\lp x_{1}|u_{1\alpha i},u_{1\beta
i},q_{i}\rp$.

Similarly generate code books
\beqa
&&U_{2\beta}^{n}(j_{2}),\quad j_{2}\in\{1,2,\cdots,2^{nR_{2\beta}}\},\\
&&U_{2\alpha}^{n}(j_{2},k_{2}),\quad k_{2}\in\{1,2,\cdots,2^{nR_{2\alpha}}\},\\
&& X_{2}^{n}(j_{2},k_{2},l_{2}),\quad l_{2}\in\{1,2,\cdots,2^{nR_{2p}}\}.\eeqa

\subsubsection*{Encoding}

Transmitter $\text{Tx}_{1}$ sends $X_{1}^{n}(j_{1},k_{1},l_{1})$ to
communicate the message indexed by $\lp j_{1},k_{1},l_{1}\rp$. Transmitter
$\text{Tx}_{2}$ sends $X_{2}^{n}(j_{2},k_{2},l_{2})$ to communicate the
message indexed by $\lp j_{2},k_{2},l_{2}\rp$.

\subsubsection*{Decoding}

The receivers do joint typical set decoding. Let $A^{\lp
n\rp}_{\epsilon}\lp \Omega \rp$ denote the set of jointly typical
sequences $\om^{n}$ where $\Omega$ is the probability space
containing the entire collection of random variables.

Receiver $\text{Rx}_{1\beta}$ determines a \emph{unique}
$\lp \hat{j}_{1},\hat{k}_{1},\hat{l}_{1}\rp$ and any $\hat{j}_{2}$ such
that
\begin{align}
\lp Q^{n},
U_{1\beta}^{n}\lp\hat{j}_{1}\rp,U_{1\alpha}^{n}\lp\hat{j}_{1},\hat{k}_{1}\rp,X_{1}^{n}\lp\hat{j}_{1},\hat{k}_{1},\hat{l}_{1}\rp,U_{2\beta}^{n}\lp\hat{j}_{2}\rp,Y_{1\beta}^{n}\rp
&\in \nn\\
A^{\lp n\rp}_{\epsilon}&\lp Q,
U_{1\beta},U_{1\alpha},X_{1},U_{2\beta},Y_{1\beta}\rp.  \nn
\end{align}
It declares an error if it fails to find such a choice.

Receiver $\text{Rx}_{1\alpha}$ determines a \emph{unique}
$\lp\hat{j}_{1},\hat{k}_{1},\hat{l}_{1}\rp$ and any
$\lp\hat{j}_{2},\hat{k}_{2}\rp$ such that,
\begin{align}
\lp Q^{n},
U_{1\beta}^{n}\lp\hat{j}_{1}\rp,U_{1\alpha}^{n}\lp\hat{j}_{1},\hat{k}_{1}\rp,X_{1}^{n}\lp\hat{j}_{1},\hat{k}_{1},\hat{l}_{1}\rp,
U_{2\beta}^{n}\lp\hat{j}_{2}\rp,
U_{2\alpha}^{n}\lp\hat{j}_{2},\hat{k}_{2}\rp, Y_{1\beta}^{n} \rp
&\in \nn\\
A^{\lp n\rp}_{\epsilon}\lp Q,
U_{1\beta},U_{1\alpha},X_{1},U_{2\beta},U_{2\alpha},Y_{1\alpha}\rp&.
\nn
\end{align}
It declares an error if it fails to find such a choice.

Similar decoding is done by receivers $\text{Rx}_{2\beta}$ and
$\text{Rx}_{2\alpha}$.

From the analysis of the probability of error, we show in
Appendix~\ref{app:Pe} that the rate vector
$(R_{1p},R_{1\alpha},R_{1\beta},R_{2p},R_{2\alpha},R_{2\beta})$ is
achievable if it satisfies the following conditions: {\footnotesize
\begin{align}
R_{1p}&\leq
I(Y_{1\beta};X_{1}|U_{1\alpha},U_{2\beta},Q),\quad \text{if } R_{1p}>0,\label{eq:1b1}\\
R_{2\beta}+R_{1p} &\leq
I(Y_{1\beta};X_{1},U_{2\beta}|U_{1\alpha},Q),\quad \text{if } R_{1p}>0,\\
R_{1\alpha}+R_{1p} &\leq
I(Y_{1\beta};X_{1}|U_{1\beta},U_{2\beta},Q),\quad \text{if } R_{1\alpha}+R_{1p}>0,\\
R_{2\beta}+R_{1\alpha}+R_{1p} &\leq
I(Y_{1\beta};X_{1},U_{2\beta}|U_{1\beta},Q),\quad \text{if } R_{1\alpha}+R_{1p}>0\\
R_{1\beta}+R_{1\alpha}+R_{1p} &\leq
I(Y_{1\beta};X_{1}|U_{2\beta},Q),\quad \text{if } R_{1\beta}+R_{1\alpha}+R_{1p}>0,\\
R_{2\beta}+R_{1\beta}+R_{1\alpha}+R_{1p} &\leq
I(Y_{1\beta};X_{1},U_{2\beta}|Q),\quad \text{if }
R_{1\beta}+R_{1\alpha}+R_{1p}>0,\label{eq:1b6}
\end{align}

\begin{align}
R_{1p} &\leq
I(Y_{1\alpha};X_{1}|U_{1\alpha},U_{2\alpha},Q),\quad \text{if } R_{1p}>0,\label{eq:1a1}\\
R_{2\alpha}+R_{1p} &\leq
I(Y_{1\alpha};X_{1},U_{2\alpha}|U_{1\alpha},U_{2\beta},Q),\quad \text{if } R_{1p}>0,\\
R_{2\beta}+R_{2\alpha}+R_{1p} &\leq
I(Y_{1\alpha};X_{1},U_{2\alpha}|U_{1\alpha},Q),\quad \text{if } R_{1p}>0,\\
R_{1\alpha}+R_{1p} &\leq
I(Y_{1\alpha};X_{1}|U_{1\beta},U_{2\alpha},Q),\quad \text{if } R_{1\alpha}+R_{1p}>0,\\
R_{2\alpha}+R_{1\alpha}+R_{1p} &\leq
I(Y_{1\alpha};X_{1},U_{2\alpha}|U_{1\beta},U_{2\beta},Q),\quad \text{if } R_{1\alpha}+R_{1p}>0,\\
R_{2\beta}+R_{2\alpha}+R_{1\alpha}+R_{1p} &\leq
I(Y_{1\alpha};X_{1},U_{2\alpha}|U_{1\beta},Q),\quad \text{if } R_{1\alpha}+R_{1p}>0,\\
R_{1\beta}+R_{1\alpha}+R_{1p} &\leq
I(Y_{1\alpha};X_{1}|U_{2\alpha},Q),\quad \text{if } R_{1\beta}+R_{1\alpha}+R_{1p}>0,\\
R_{2\alpha}+R_{1\beta}+R_{1\alpha}+R_{1p} &\leq
I(Y_{1\alpha};X_{1},U_{2\alpha}|U_{2\beta},Q),\quad \text{if } R_{1\beta}+R_{1\alpha}+R_{1p}>0,\\
R_{2\beta}+R_{2\alpha}+R_{1\beta}+R_{1\alpha}+R_{1p} &\leq
I(Y_{1\alpha};X_{1},U_{2\alpha}|Q),\quad \text{if }
R_{1\beta}+R_{1\alpha}+R_{1p}>0,\label{eq:1a9}
\end{align}

\begin{align}
R_{2p}&\leq
I(Y_{2\beta};X_{2}|U_{2\alpha},U_{1\beta},Q),\quad \text{if } R_{2p}>0,\label{eq:2b1}\\
R_{1\beta}+R_{2p} &\leq
I(Y_{2\beta};X_{2},U_{1\beta}|U_{2\alpha},Q),\quad \text{if } R_{2p}>0,\\
R_{2\alpha}+R_{2p} &\leq
I(Y_{2\beta};X_{2}|U_{2\beta},U_{1\beta},Q),\quad \text{if } R_{2\alpha}+R_{2p}>0,\\
R_{1\beta}+R_{2\alpha}+R_{2p} &\leq
I(Y_{2\beta};X_{2},U_{1\beta}|U_{2\beta},Q),\quad \text{if } R_{2\alpha}+R_{2p}>0,\\
R_{2\beta}+R_{2\alpha}+R_{2p} &\leq
I(Y_{2\beta};X_{2}|U_{1\beta},Q),\quad \text{if } R_{2\beta}+R_{2\alpha}+R_{2p}>0,\\
R_{1\beta}+R_{2\beta}+R_{2\alpha}+R_{2p} &\leq
I(Y_{2\beta};X_{2},U_{1\beta}|Q),\quad \text{if }
R_{2\beta}+R_{2\alpha}+R_{2p}>0,\label{eq:2b6}
\end{align}

\begin{align}
R_{2p} &\leq
I(Y_{2\alpha};X_{2}|U_{2\alpha},U_{1\alpha},Q),\quad \text{if } R_{2p}>0,\label{eq:2a1}\\
R_{1\alpha}+R_{2p} &\leq
I(Y_{2\alpha};X_{2},U_{1\alpha}|U_{2\alpha},U_{1\beta},Q),\quad \text{if } R_{2p}>0,\\
R_{1\beta}+R_{1\alpha}+R_{2p} &\leq
I(Y_{2\alpha};X_{2},U_{1\alpha}|U_{2\alpha},Q),\quad \text{if } R_{2p}>0,\\
R_{2\alpha}+R_{2p} &\leq
I(Y_{2\alpha};X_{2}|U_{2\beta},U_{1\alpha},Q),\quad \text{if } R_{2\alpha}+R_{2p}>0,\\
R_{1\alpha}+R_{2\alpha}+R_{2p} &\leq
I(Y_{2\alpha};X_{2},U_{1\alpha}|U_{2\beta},U_{1\beta},Q),\quad \text{if } R_{2\alpha}+R_{2p}>0,\\
R_{1\beta}+R_{1\alpha}+R_{2\alpha}+R_{2p} &\leq
I(Y_{2\alpha};X_{2},U_{1\alpha}|U_{2\beta},Q),\quad \text{if } R_{2\alpha}+R_{2p}>0,\\
R_{2\beta}+R_{2\alpha}+R_{2p} &\leq
I(Y_{2\alpha};X_{2}|U_{1\alpha},Q),\quad \text{if } R_{2\beta}+R_{2\alpha}+R_{2p}>0,\\
R_{1\alpha}+R_{2\beta}+R_{2\alpha}+R_{2p} &\leq
I(Y_{2\alpha};X_{2},U_{1\alpha}|U_{1\beta},Q),\quad \text{if } R_{2\beta}+R_{2\alpha}+R_{2p}>0,\\
R_{1\beta}+R_{1\alpha}+R_{2\beta}+R_{2\alpha}+R_{2p} &\leq
I(Y_{2\alpha};X_{2},U_{1\alpha}|Q),\quad \text{if }
R_{2\beta}+R_{2\alpha}+R_{2p}>0,\label{eq:2a9}
\end{align}

\begin{align}
R_{1p}\geq0,\label{eq:nn1}\\
R_{1\alpha}\geq0,\\
R_{1\beta}\geq0,\\
R_{2p}\geq0,\\
R_{2\alpha}\geq0,\\
R_{2\beta}\geq0\label{eq:nn6}.
\end{align}
}

Note that \eqref{eq:1b1}-\eqref{eq:1b6} are the decodability
conditions at $\text{Rx}_{1\beta}$; \eqref{eq:1a1}-\eqref{eq:1a9}
are the decodability conditions at $\text{Rx}_{1\alpha}$;
\eqref{eq:2b1}-\eqref{eq:2b6} are the decodability conditions at
$\text{Rx}_{2\beta}$; \eqref{eq:2a1}-\eqref{eq:2a9} are the
decodability conditions at $\text{Rx}_{2\alpha}$ and
\eqref{eq:nn1}-\eqref{eq:nn6} are stating the fact that the rates
are nonnegative real numbers.

Define \beq \tilde{{\mathcal R}}^{(6)}_{in}(P) \df \lbr
(R_{1p},R_{1\alpha},R_{1\beta},R_{2p},R_{2\alpha},R_{2\beta}):
\text{satisfies \eqref{eq:1b1}-\eqref{eq:nn6}} \label{eq:tRin6}\rbr,
\eeq and its projection onto the two dimension space $(R_{1},R_{2})$
by $\tilde{{\mathcal R}}_{in}(P)$.

\begin{lem}\label{lem:ib}
\[{\mathcal R}_{in}(P) \subseteq \tilde{{\mathcal R}}_{in}(P).\]
\end{lem}
\noindent\emph{Proof:} See Appendix~\ref{app:Rp}. \hfill$\Box$

Thus, we have shown that the capacity region ${\mathcal C}$
satisfies
 \beq {\mathcal C} \supseteq \bigcup_{P\in{\mathcal P}} {\mathcal
R}_{in}\lp P \rp.\eeq In particular, restricting to a subfamily
of ${\mathcal P}$, where given random variables $(Q,X_1,X_2)$ such
that $X_1 - Q - X_2$ is a Markov chain and
$(U_{1\alpha},U_{1\beta},U_{2\alpha},U_{2\beta})$ are defined
by~\eqref{eq:aux}, we get \beq {\mathcal C} \supseteq \bigcup_{\lp
Q,X_1,X_2 \rp} {\mathcal R}_{in}\lp Q,X_1,X_2 \rp.\eeq
This completes the proof of Theorem~\ref{thm:Main1}. \hfill$\Box$

\subsection{Geometric Properties Of ${\mathcal R}_{in}\lp Q,X_{1},X_{2}\rp $} \label{sec:extRep}
We have noted earlier that it is tedious to characterize ${\mathcal
R}_{in}(Q,X_{1},X_{2})$ explicitly. Nevertheless, we would like to derive some
useful insights into the geometric properties of  ${\mathcal
R}_{in}\lp Q,X_{1},X_{2}\rp$. These will prove useful in deriving the
outer bound.

We begin by noting that ${\mathcal R}_{in}\lp Q,X_{1},X_{2}\rp$ is a closed and bounded
convex region. (In fact, we know that it is a polyhedron.) The
\emph{extremal representation theorem} of classical Convex set
theory (see Theorem 18.8, \cite{ConAnal}) states that ``an
$n$-dimensional closed convex set in $\real^{n}$ is the intersection
of the closed half-spaces tangent to it''. Thus, \beq {\mathcal
R}_{in}(Q,X_{1},X_{2}) = \lbr (R_{1},R_{2}): aR_{1}+bR_{2} \leq
c^{*}\lp a,b|(Q,X_{1},X_{2})\rp,\;\forall(a,b)\in\real^{2} \rbr.
\label{eq:extRep}\eeq Here, $c^{*}\lp a,b|(Q,X_{1},X_{2}) \rp$ is
the support function (Section 13, \cite{ConAnal}) of ${\mathcal
R}_{in}(Q,X_{1},X_{2})$ and is defined as the solution of the
following linear program,
\begin{align}
\mathrm{Max}&\quad aR_{1}+bR_{2}, \label{eq:primal1} \\
\nn\\
\mathrm{s.t.}&\quad (R_{1},R_{2})\in R_{in}(Q,X_{1},X_{2}). \nn
\end{align}
Since ${\mathcal R}_{in}\lp Q,X_{1},X_{2}\rp$ is the projection of the
six-dimensional region ${\mathcal R}^{(6)}_{in}\lp Q,X_{1},X_{2}\rp$, the
linear program~\eqref{eq:primal1} is equivalent to the following
linear program.
\begin{align}
\mathrm{Max}&\quad aR_{1p}+aR_{1\alpha}+aR_{1\beta}+bR_{2p}+bR_{2\alpha}+bR_{2\beta},
\label{eq:primal2} \\
\nn\\
\mathrm{s.t.}&\quad (R_{1p}, R_{1\alpha}, R_{1\beta}, R_{2p},
R_{2\alpha},R_{2\beta})\in R_{in}^{(6)}(Q,X_{1},X_{2}). \nn
\end{align}

The {\em dual} of the linear program in Equation~\eqref{eq:primal2} sheds important geometric
information. Let us denote the dual-variables associated with the
inequalities (\ref{ibeq1b1})-(\ref{ibeq1b6}) by
$\nu_{11},\ldots,\nu_{16}$, with (\ref{ibeq1a1})-(\ref{ibeq1a9}) by
$\mu_{11},\ldots,\mu_{19}$, with (\ref{ibeq2b1})-(\ref{ibeq2b6}) by
$\nu_{21},\ldots,\nu_{26}$, with (\ref{ibeq2a1})-(\ref{ibeq2a9}) by
$\mu_{21},\ldots,\mu_{29}$ and with (\ref{ibeqnn1})-(\ref{ibeqnn2})
by $\om_{1}$ and $\om_{2}$.

Define $\La_{(a,b)} \subset \real^{32}$ by,
\begin{align}
\La_{(a,b)} \df \lbr \left(\{{\nu}_{1i}\}_{1}^{6},
\{{\mu}_{1i}\}_{1}^{9}, \{{\nu}_{2i}\}_{1}^{6},
\{{\mu}_{2i}\}_{1}^{9},\{\om_{i}\}_{1}^{2}\right):
\mathrm{satisfying~\eqref{eq:dual1}-\eqref{eq:dual7}} \rbr.
\label{eq:La}
\end{align}

\begin{eqnarray}
\sum_{i=1}^{6}\nu_{1i}+\sum_{i=1}^{9}\mu_{1i}-\om_{1}&=&a\label{eq:dual1}\\
\sum_{i=3}^{6}\nu_{1i}+\sum_{i=4}^{9}\mu_{1i} + (\mu_{22}+\mu_{25}+\mu_{28})+(\mu_{23}+\mu_{26}+\mu_{29})-\om_{1}&=&a \label{eq:dual2} \\
\sum_{i=5}^{6}\nu_{1i}+\sum_{i=7}^{9}\mu_{1i}+(\mu_{23}+\mu_{26}+\mu_{29})+(\nu_{22}+\nu_{24}+\nu_{26})-\om_{1}&=&a  \label{eq:dual3}\\
\sum_{i=1}^{6}\nu_{2i}+\sum_{i=1}^{9}\mu_{2i}-\om_{2}&=&b \label{eq:dual4}\\
\sum_{i=3}^{6}\nu_{2i}+\sum_{i=4}^{9}\mu_{2i} + (\mu_{12}+\mu_{15}+\mu_{18})+(\mu_{13}+\mu_{16}+\mu_{19})-\om_{2}&=&b \label{eq:dual5}\\
\sum_{i=5}^{6}\nu_{2i}+\sum_{i=7}^{9}\mu_{2i}+(\mu_{13}+\mu_{16}+\mu_{19})+(\nu_{12}+\nu_{14}+\nu_{16})-\om_{2}&=&b \label{eq:dual6}\\
\mu_{ij},\nu_{ij},\om_{i}&\geq&0. \label{eq:dual7}
\end{eqnarray}
For any $\la \in \La_{(a,b)}$ define \beq
c^{(in)}_{(\la,a,b)}(Q,X_{1},X_{2}) \df
\sum_{j=1}^2\left(\sum_{i=1}^6 \nu_{ji}\ga_{ji}+\sum_{i=1}^9
\mu_{ji}\de_{ji}\right). \eeq The dual linear program is
\begin{align}
\mathrm{min}&\quad c^{(in)}_{(\la,a,b)}(Q,X_{1},X_{2}), \\
\nn\\
\mathrm{such that}&\quad \la \in \La_{(a,b)}. \nn
\end{align}
By the strong duality theorem
\begin{align}
c^{*}\lp a,b|(Q,X_{1},X_{2}) \rp = \mathrm{min} \lbr
c^{(in)}_{(\la,a,b)}(Q,X_{1},X_{2})|\la\in\La_{(a,b)} \rbr.
\end{align}
Therefore,
\begin{align}
\lbr (R_{1},R_{2}):aR_{1}+bR_{2} \leq c^{*}\lp a,b|(Q,X_{1},X_{2})\rp \rbr &= \nn \\
\bigcap_{\la\in\La_{(a,b)}} & \lbr (R_{1},R_{2}):aR_{1}+bR_{2} \leq
c^{(in)}_{(\la,a,b)}(Q,X_{1},X_{2}) \rbr. \label{eq:supPlane}
\end{align}

Using~\eqref{eq:extRep} and~\eqref{eq:supPlane}, $\mathcal
R_{in}(Q,X_{1},X_{2})$ can now be described as,
\begin{align}
\mathcal R_{in}(Q,X_{1},X_{2})
= \{(R_{1},R_{2}):\;\;&aR_{1}+bR_{2} \leq c^{(in)}_{(\la,a,b)}(Q,X_{1},X_{2}),\nonumber \\
& \forall \la \in \La_{(a,b)},\quad \forall\;(a,b)\in\real^{2}\}.
\label{eq:Rinbig}
\end{align}

The set of linear inequalities (c.f.\ \eqref{eq:Rinbig}) that is used to
describe $\mathcal R_{in}(Q,X_{1},X_{2})$ is very large and many of
the inequalities might be redundant. The following result, Lemma~\ref{lem:RinProp}, tries
to characterize some of these redundant inequalities.

Let $\La_{(a,b)}^{\prime}$ be a subset of $\La_{(a,b)}$ defined by
\beq \La_{(a,b)}^{\prime} \df \lbr \la \in \La_{(a,b)}:
\om_{1}=0,\;\om_{2}=0 \rbr. \eeq

\begin{lem}\label{lem:RinProp}
\begin{align}
\mathcal R_{in}(Q,X_{1},X_{2})
= \{(R_{1},R_{2})&:\;R_{1}\geq0,\;R_{2}\geq0 \nn\\
&aR_{1}+bR_{2} \leq c^{(in)}_{(\la,a,b)}(Q,X_{1},X_{2}),\; \forall
\la \in \La_{(a,b)}^{\prime},\;\forall a\geq0,\;b\geq0\}.
\label{eq:Rin}
\end{align}
\end{lem}
\emph{Proof:} Every inequality used to define $\mathcal
R_{in}(Q,X_{1},X_{2})$ in~\eqref{eq:Rinbig} is described by
parameters $(a,b)$ and $\la \in \La_{(a,b)}$. Note that this set of
inequalities includes the following two inequalities:
\begin{align}
-R_{1}&\leq0 \\
-R_{2}&\leq0.
\end{align}

Consider any inequality, other than the two special ones above,
described by $(a,b)$ and  $\la \in \La_{(a,b)}$, such that
$\la\not\in\La_{(a,b)}^{\prime}$: \beq aR_{1}+bR_{2} \leq
c^{(in)}_{(\la,a,b)}. \label{eq:redineq} \eeq Define
\beq(\tilde{a},\tilde{b})\df(a+\om_{1},b+\om_{2}).\eeq Consider
$\tilde\la\in\La_{(\tilde{a},\tilde{b})}^{\prime}$, obtained by replacing
$\om_{1}$ and $\om_{2}$ in $\la$ by $0$. Now we have
\beq
c^{(in)}_{(\la,a,b)} = c^{(in)}_{(\tilde\la,\tilde{a},\tilde{b})}. \eeq
Therefore \beq (a+\om_{1}) R_{1} + (b+\om_{2}) R_{2} =
\tilde{a}R_{1}+\tilde{b}R_{2}\leq
c^{(in)}_{(\tilde\la,\tilde{a},\tilde{b})} = c^{(in)}_{(\la,a,b)}.
\label{eq:greenineq} \eeq The above inequality, along with $R_{1}
\geq 0$ and $R_{2} \geq 0$, implies~\eqref{eq:redineq}. Therefore we
have that \eqref{eq:redineq} is redundant.

Thus we have proved that inequalities that are characterized by a
$\la\not\in\La_{(a,b)}^{\prime}$ are redundant and can be removed.
It also follows from~\eqref{eq:dual1}-\eqref{eq:dual7} that
$\La_{(a,b)}^{\prime}$ is an empty set if either $a$ or $b$ is less
than 0. Thus we only need to consider inequalities characterized by
$(a,b)$, where $a\geq0$ and $b\geq0$. This completes the proof.
\hfill$\Box$

We also state the following proposition, that will be used in
proving the outer bound.

\begin{prop}\label{prop:dualsol}
For any $\la \in\La_{(a,b)}^{\prime}$,
\begin{align}
(\mu_{11}+\mu_{12}+\mu_{13})+(\nu_{11}+\nu_{12})
 &= (\mu_{22}+\mu_{23}+\mu_{25}+\mu_{26}+\mu_{28}+\mu_{29}) \label{eq:dualsol1} \\
(\mu_{14}+\mu_{15}+\mu_{16})+(\mu_{22}+\mu_{25}+\mu_{28})+(\nu_{13}+\nu_{14})
 &= (\nu_{22}+\nu_{24}+\nu_{26}) \label{eq:dualsol2}\\
(\mu_{21}+\mu_{22}+\mu_{23})+(\nu_{21}+\nu_{22})
 &= (\mu_{12}+\mu_{13}+\mu_{15}+\mu_{16}+\mu_{18}+\mu_{19})\label{eq:dualsol3} \\
(\mu_{24}+\mu_{25}+\mu_{26})+(\mu_{12}+\mu_{15}+\mu_{18})+(\nu_{23}+\nu_{24})
 &= (\nu_{12}+\nu_{14}+\nu_{16}).  \label{eq:dualsol4}
\end{align}
\end{prop}
\emph{Proof:} The proof follows directly
from~\eqref{eq:dual1}-\eqref{eq:dual6}. \hfill$\Box$

\section{Outer Bound}
\label{sec:ob}

Our goal in this Section is to show that, if $(R_{1},R_{2})$ is
achievable then there exist random variables $({Q},{X}_1,{X}_2)$,
where ${X}_1 - {Q} - {X}_2$ is a Markov chain, and a region
\begin{align}
\mathcal R_{out}({Q},{X}_{1},{X}_{2})
\df \{(R_{1},R_{2})&:\;R_{1}\geq0,\;R_{2}\geq0 \nn\\
&aR_{1}+bR_{2} \leq c^{(out)}_{(\la,a,b)}({Q},{X}_{1},{X}_{2}),
\;\forall \la \in \La_{(a,b)}^{\prime},\;\forall a\geq0,\;b\geq0\},
\label{eq:Rout}
\end{align}
such that \beq
(R_{1},R_{2}) \in \mathcal
R_{out}({Q},{X}_{1},{X}_{2}).
\eeq
The term $c^{(out)}_{(\la,a,b)}({Q},{X}_{1},{X}_{2})$ is defined in
Section~\ref{sec:thm2-1}. Note that our definition of $\mathcal
R_{out}({Q},{X}_{1},{X}_{2})$ is inspired by the characterization of
$\mathcal R_{in}(Q,X_{1},X_{2})$  that we have obtained through
Lemma~\ref{lem:RinProp}.

Further, quantifying the difference between
$c^{(in)}_{(\la,a,b)}({Q},{X}_{1},{X}_{2})$ and
$c^{(out)}_{(\la,a,b)}({Q},{X}_{1},{X}_{2})$ will give us the gap
between the inner and the outer bounds.

\subsection{Proof Of Theorem~\ref{thm:Main2} (i)}\label{sec:thm2-1}

Suppose  there is a sequence of encoders at  rates
$(R_1,R_2)$, sequenced by the block length $n$, and decoders with probability of
error going to $0$ as $n\rightarrow\infty$. Fix the block length $n$ and consider
the  corresponding code book.   Let
$X_1^n,X_2^n,S_1^n,S_2^n,Y_1^n,Y_2^n$ be the random variables
induced by the channel and encoders for uniformly distributed
messages, independent across the two users. We define random
variables $U_{1\alpha}^n$ which is obtained by passing $X_1^n$
through an independent copy of the channel $p_{S_{1\alpha}|X_1}$,
and $U_{1\beta}^n$ by passing the $U_{1\alpha}^n$ so obtained
through an independent copy of rhe channel
$p_{S_{1\beta}|S_{1\alpha}}$.  Similarly, we also define
$U_{2\alpha}^n$ and $U_{2\beta}^n$ from $X_2^n$ and independent
copies of $p_{S_{2\alpha}|X_2}$ and $p_{S_{2\beta}|S_{2\alpha}}$.

Consider any non-negative pair $(a,b)$ and any $\la \in
\La_{(a,b)}^{\prime}$. Since the probability of error goes to $0$ as
$n\rightarrow\infty$, by Fano's inequality there exists a sequence
$\epsilon_{n}\rightarrow0$ such that
\begin{align}
&n(aR_1+bR_2-(a+b)\epsilon_n)\nonumber\\
&\stackrel{\text(1)}{\leq}
\left(\sum_{i=1}^6{\nu}_{1i}\right)I(X_1^n;Y_{1\beta}^n)+
\left(\sum_{i=1}^9 {\mu}_{1i}\right)I(X_1^n;Y_{1\alpha}^n)\nonumber\\
&\;\;+ \left(\sum_{i=1}^6{\nu}_{2i}\right)I(X_2^n;Y_{2\beta}^n)+
\left(\sum_{i=1}^9 {\mu}_{2i}\right)I(X_2^n;Y_{2\alpha}^n)\nonumber\\
&\stackrel{\text(2)}{\leq} \sum_{i=1}^6
{\nu}_{1i}I(X_1^n;Y_{1\beta}^n,V_{1i\beta}^{n})+
\sum_{i=1}^9 {\mu}_{1i}I(X_1^n;Y_{1\alpha}^n,V_{1i\alpha}^{n})\nonumber\\
&\;\; + \sum_{i=1}^6
{\nu}_{2i}I(X_2^n;Y_{2\beta}^n,V_{2i\beta}^{n})+ \sum_{i=1}^9
{\mu}_{2i}I(X_2^n;Y_{2\alpha}^n,V_{2i\alpha}^{n}).
\label{eq:upperbound}
\end{align}

Note that in step (1), we split up $a$ and $b$ according to
\eqref{eq:dual1} and \eqref{eq:dual4}, considering decoders under
different states. In step (2), we consider {\em genies} which
provide different side-information $V$'s to the decoders. Consider,
for instance, the term
${\nu}_{11}I(X_1^n;Y_{1\beta}^n,V_{11\beta}^{n})$. We will choose
the side-information $V_{11\beta}^{n}$ in such a way that we can
form a correspondence between this term and the term contributed to
the inner bound by the right hand side of the constraint
\eqref{ibeq1b1}. In particular, we choose the genie provided
side-information $V_{11\beta}^{n}$ to match the error-event
corresponding to \eqref{ibeq1b1}. More specifically, we note that
the corresponding error-event is when receiver~1 in state~$\beta$
correctly decodes the other user's common information $U_{2\beta}$,
and its own common information $(U_{1\beta},U_{1\alpha})$, but makes
an error in decoding its private message. Hence, the genie provides
the side-information $(U_{1\alpha}^n,U_{1\beta}^n,U_{2\beta}^n)$
which can be shrunk to
$V_{11\beta}^{n}=(U_{1\alpha}^n,U_{2\beta}^n)$ because of the Markov
relationship between $X_1^n, U_{1\alpha}^{n}$, and $U_{1\beta}^{n}$.
Now, we expand the term $I(X_1^n;Y_{1\beta}^n,V_{11\beta}^{n})$
to get \eqref{obeq1b1}. We can repeat these two steps for every term in
\eqref{eq:upperbound}: the (expanded) upper bounds on all the terms are given in
\eqref{obeq1b1}-\eqref{obeq2a9}.

{\footnotesize
\begin{align}
I(X_1^n;Y_{1\beta}^n,V_{11\beta}^{n}) &\df
I(X_1^n;Y_{1\beta}^n,U_{1\alpha}^n,U_{2\beta}^n)&=&\;
H(Y_{1\beta}^n|U_{1\alpha}^n,U_{2\beta}^n)-H(S_{2\beta}^n|U_{2\beta}^n)+H(U_{1\alpha}^n)-H(U_{1\alpha}^n|X_1^n) \label{obeq1b1}\\
I(X_1^n;Y_{1\beta}^n,V_{12\beta}^{n}) &\df
I(X_1^n;Y_{1\beta}^n,U_{1\alpha}^n) &=&\;
H(Y_{1\beta}^n|U_{1\alpha}^n)-H(S_{2\beta}^n)+H(U_{1\alpha}^n)-H(U_{1\alpha}^n|X_1^n)\\
I(X_1^n;Y_{1\beta}^n,V_{13\beta}^{n}) &\df
I(X_1^n;Y_{1\beta}^n,U_{1\beta}^n,U_{2\beta}^n) &=&\;
H(Y_{1\beta}^n|U_{1\beta}^n,U_{2\beta}^n)-H(S_{2\beta}^n|U_{2\beta}^n)+H(U_{1\beta}^n)-H(U_{1\beta}^n|X^n_1) \\
I(X_1^n;Y_{1\beta}^n,V_{14\beta}^{n}) &\df
I(X_1^n;Y_{1\beta}^n,U_{1\beta}^n) &=&\;
H(Y_{1\beta}^n|U_{1\beta}^n)-H(S_{2\beta}^n)+H(U_{1\beta}^n)-H(U_{1\beta}^n|X^n_1) \\
I(X_1^n;Y_{1\beta}^n,V_{15\beta}^{n}) &\df
I(X_1^n;Y_{1\beta}^n,U_{2\beta}^n) &=&\;
H(Y_{1\beta}^n|U_{2\beta}^n)-H(S_{2\beta}^n|U_{2\beta}^n) \\
I(X_1^n;Y_{1\beta}^n,V_{16\beta}^{n}) &\df
I(X_1^n;Y_{1\beta}^n) &=&\; H(Y_{1\beta}^n)-H(S_{2\beta}^n)
\label{obeq1b6}
\end{align}

\begin{align}
I(X_1^n;Y_{1\alpha}^n,V_{11\alpha}^{n}) &\df
I(X_1^n;Y_{1\alpha}^n,U_{1\alpha}^n,U_{2\alpha}^n) &=&\;
H(Y_{1\alpha}^n|U_{1\alpha}^n,U_{2\alpha}^n)-H(S_{2\alpha}^n|U_{2\alpha}^n)+H(U_{1\alpha}^n)-H(U_{1\alpha}^n|X_1^n) \label{obeq1a1}\\
I(X_1^n;Y_{1\alpha}^n,V_{12\alpha}^{n}) &\df
I(X_1^n;Y_{1\alpha}^n,U_{1\alpha}^n,U_{2\beta}^n) &=&\;
H(Y_{1\alpha}^n|U_{1\alpha}^n,U_{2\beta}^n)-H(S_{2\alpha}^n|U_{2\beta}^n)+H(U_{1\alpha}^n)-H(U_{1\alpha}^n|X_1^n) \\
I(X_1^n;Y_{1\alpha}^n,V_{13\alpha}^{n}) &\df
I(X_1^n;Y_{1\alpha}^n,U_{1\alpha}^n) &=&\;
H(Y_{1\alpha}^n|U_{1\alpha}^n)-H(S_{2\alpha}^n)+H(U_{1\alpha}^n)-H(U_{1\alpha}^n|X_1^n) \\
I(X_1^n;Y_{1\alpha}^n,V_{14\alpha}^{n}) &\df
I(X_1^n;Y_{1\alpha}^n,U_{1\beta}^n,U_{2\alpha}^n) &=&\;
H(Y_{1\alpha}^n|U_{1\beta}^n,U_{2\alpha}^n)-H(S_{2\alpha}^n|U_{2\alpha}^n)+H(U_{1\beta}^n)-H(U_{1\beta}^n|X^n_1) \\
I(X_1^n;Y_{1\alpha}^n,V_{15\alpha}^{n}) &\df
I(X_1^n;Y_{1\alpha}^n,U_{1\beta}^n,U_{2\beta}^n) &=&\;
H(Y_{1\alpha}^n|U_{1\beta}^n,U_{2\beta}^n)-H(S_{2\alpha}^n|U_{2\beta}^n)+H(U_{1\beta}^n)-H(U_{1\beta}^n|X^n_1) \\
I(X_1^n;Y_{1\alpha}^n,V_{16\alpha}^{n}) &\df
I(X_1^n;Y_{1\alpha}^n,U_{1\beta}^n) &=&\;
H(Y_{1\alpha}^n|U_{1\beta}^n)-H(S_{2\alpha}^n)+H(U_{1\beta}^n)-H(U_{1\beta}^n|X^n_1) \\
I(X_1^n;Y_{1\alpha}^n,V_{17\alpha}^{n}) &\df
I(X_1^n;Y_{1\alpha}^n,U_{2\alpha}^n) &=&\;
H(Y_{1\alpha}^n|U_{2\alpha}^n)-H(S_{2\alpha}^n|U_{2\alpha}^n) \\
I(X_1^n;Y_{1\alpha}^n,V_{18\alpha}^{n}) &\df
I(X_1^n;Y_{1\alpha}^n,U_{2\beta}^n) &=&\;
H(Y_{1\alpha}^n|U_{2\beta}^n)-H(S_{2\alpha}^n|U_{2\beta}^n) \\
I(X_1^n;Y_{1\alpha}^n,V_{19\alpha}^{n}) &\df
I(X_1^n;Y_{1\alpha}^n) &=&\; H(Y_{1\alpha}^n)-H(S_{2\alpha}^n)
\label{obeq1a9}
\end{align}

\begin{align}
I(X_2^n;Y_{2\beta}^n,V_{21\beta}^{n}) &\df
I(X_2^n;Y_{2\beta}^n,U_{2\alpha}^n,U_{1\beta}^n) &=&\;
H(Y_{2\beta}^n|U_{2\alpha}^n,U_{1\beta}^n)-H(S_{1\beta}^n|U_{1\beta}^n)+H(U_{2\alpha}^n)-H(U_{2\alpha}^n|X_2^n) \label{obeq2b1}\\
I(X_2^n;Y_{2\beta}^n,V_{22\beta}^{n}) &\df
I(X_2^n;Y_{2\beta}^n,U_{2\alpha}^n) &=&\;
H(Y_{2\beta}^n|U_{2\alpha}^n)-H(S_{1\beta}^n)+H(U_{2\alpha}^n)-H(U_{2\alpha}^n|X^2_n)\\
I(X_2^n;Y_{2\beta}^n,V_{23\beta}^{n}) &\df
I(X_2^n;Y_{2\beta}^n,U_{2\beta}^n,U_{1\beta}^n) &=&\;
H(Y_{2\beta}^n|U_{2\beta}^n,U_{1\beta}^n)-H(S_{1\beta}^n|U_{1\beta}^n)+H(U_{2\beta}^n)-H(U_{2\beta}^n|X^n_2) \\
I(X_2^n;Y_{2\beta}^n,V_{24\beta}^{n}) &\df
I(X_2^n;Y_{2\beta}^n,U_{2\beta}^n) &=&\;
H(Y_{2\beta}^n|U_{2\beta}^n)-H(S_{1\beta}^n)+H(U_{2\beta}^n)-H(U_{2\beta}^n|X^n_2) \\
I(X_2^n;Y_{2\beta}^n,V_{25\beta}^{n}) &\df
I(X_2^n;Y_{2\beta}^n,U_{1\beta}^n) &=&\;
H(Y_{2\beta}^n|U_{1\beta}^n)-H(S_{1\beta}^n|U_{1\beta}^n) \\
I(X_2^n;Y_{2\beta}^n,V_{26\beta}^{n}) &\df
I(X_2^n;Y_{2\beta}^n) &=&\; H(Y_{2\beta}^n)-H(S_{1\beta}^n)
\label{obeq2b6}
\end{align}

\begin{align}
I(X_2^n;Y_{2\alpha}^n,V_{21\alpha}^{n}) &\df
I(X_2^n;Y_{2\alpha}^n,U_{2\alpha}^n,U_{1\alpha}^n) &=&\;
H(Y_{2\alpha}^n|U_{2\alpha}^n,U_{1\alpha}^n)-H(S_{1\alpha}^n|U_{1\alpha}^n)+H(U_{2\alpha}^n)-H(U_{2\alpha}^n|X_2^n) \label{obeq2a1}\\
I(X_2^n;Y_{2\alpha}^n,V_{22\alpha}^{n}) &\df
I(X_2^n;Y_{2\alpha}^n,U_{2\alpha}^n,U_{1\beta}^n) &=&\;
H(Y_{2\alpha}^n|U_{2\alpha}^n,U_{1\beta}^n)-H(S_{1\alpha}^n|U_{1\beta}^n)+H(U_{2\alpha}^n)-H(U_{2\alpha}^n|X_2^n) \\
I(X_2^n;Y_{2\alpha}^n,V_{23\alpha}^{n}) &\df
I(X_2^n;Y_{2\alpha}^n,U_{2\alpha}^n) &=&\;
H(Y_{2\alpha}^n|U_{2\alpha}^n)-H(S_{1\alpha}^n)+H(U_{2\alpha}^n)-H(U_{2\alpha}^n|X_2^n) \\
I(X_2^n;Y_{2\alpha}^n,V_{24\alpha}^{n}) &\df
I(X_2^n;Y_{2\alpha}^n,U_{2\beta}^n,U_{1\alpha}^n) &=&\;
H(Y_{2\alpha}^n|U_{2\beta}^n,U_{1\alpha}^n)-H(S_{1\alpha}^n|U_{1\alpha}^n)+H(U_{2\beta}^n)-H(U_{2\beta}^n|X^n_2) \\
I(X_2^n;Y_{2\alpha}^n,V_{25\alpha}^{n}) &\df
I(X_2^n;Y_{2\alpha}^n,U_{2\beta}^n,U_{1\beta}^n) &=&\;
H(Y_{2\alpha}^n|U_{2\beta}^n,U_{1\beta}^n)-H(S_{1\alpha}^n|U_{1\beta}^n)+H(U_{2\beta}^n)-H(U_{2\beta}^n|X^n_2) \\
I(X_2^n;Y_{2\alpha}^n,V_{26\alpha}^{n}) &\df
I(X_2^n;Y_{2\alpha}^n,U_{2\beta}^n) &=&\;
H(Y_{2\alpha}^n|U_{2\beta}^n)-H(S_{1\alpha}^n)+H(U_{2\beta}^n)-H(U_{2\beta}^n|X^n_2) \\
I(X_2^n;Y_{2\alpha}^n,V_{27\alpha}^{n}) &\df
I(X_2^n;Y_{2\alpha}^n,U_{1\alpha}^n) &=&\;
H(Y_{2\alpha}^n|U_{1\alpha}^n)-H(S_{1\alpha}^n|U_{1\alpha}^n) \\
I(X_2^n;Y_{2\alpha}^n,V_{28\alpha}^{n}) &\df
I(X_2^n;Y_{2\alpha}^n,U_{1\beta}^n) &=&\;
H(Y_{2\alpha}^n|U_{1\beta}^n)-H(S_{1\alpha}^n|U_{1\beta}^n)\\
I(X_2^n;Y_{2\alpha}^n,V_{29\alpha}^{n}) &\df
I(X_2^n;Y_{2\alpha}^n) &=&\; H(Y_{2\alpha}^n)-H(S_{1\alpha}^n).
\label{obeq2a9}
\end{align}}
Continuing with our outer bound derivation, from \eqref{eq:upperbound},
\begin{align}
& n(aR_{1}+bR_{2}-(a+b)\epsilon_{n}) \nonumber \\
&\stackrel{\text{(a)}}{\leq} \sum_{i=1}^6
{\nu}_{1i}H(Y_{1\beta}^n|V_{1i\beta}^{n})+
\sum_{i=1}^9 {\mu}_{1i}H(Y_{1\alpha}^n|V_{1i\alpha}^{n}) \nonumber \\
&\;\; +
H(U_{1\alpha}^{n})\{({\nu}_{11}+{\nu}_{12})+({\mu}_{11}+{\mu}_{12}+{\mu}_{13})\}\;+\;
H(U_{1\beta}^{n})\{(\nu_{13}+\nu_{14})+(\mu_{14}+\mu_{15}+\mu_{16})\}\nonumber \\
&\;\; - H(S_{1\alpha}^{n})\{({\mu}_{23}+\mu_{26}+\mu_{29})\}\;-\;
H(S_{1\beta}^{n})\{({\nu}_{22}+\nu_{24}+\nu_{26})\}\nonumber \\
&\;\; -
H(U_{1\alpha}^{n}|X_{1}^{n})\{({\nu}_{11}+{\nu}_{12})+({\mu}_{11}+{\mu}_{12}+{\mu}_{13})\}\nn\\
&\;\; -
H(U_{1\beta}^{n}|X_{1}^{n})\{(\nu_{13}+\nu_{14})+(\mu_{14}+\mu_{15}+\mu_{16})\}\nonumber \\
&\;\; -
H(S_{1\alpha}^{n}|U_{1\alpha}^{n})\{({\mu}_{21}+\mu_{24}+\mu_{27})\}\;-\;
H(S_{1\beta}^{n}|U_{1\beta}^{n})\{({\nu}_{21}+\nu_{23}+\nu_{25})\}\nonumber \\
&\;\; - H(S_{1\alpha}^{n}|U_{1\beta}^{n})\{({\mu}_{22}+\mu_{25}+\mu_{28})\} \nn \\
&\;\;+ \sum_{i=1}^6 {\nu}_{2i}H(Y_{2\beta}^n|V_{2i\beta}^{n})+
\sum_{i=1}^9 {\mu}_{2i}H(Y_{2\alpha}^n|V_{2i\alpha}^{n}) \nonumber \\
&\;\; +
H(U_{2\alpha}^{n})\{({\nu}_{21}+{\nu}_{22})+({\mu}_{21}+{\mu}_{22}+{\mu}_{23})\}\;+\;
H(U_{2\beta}^{n})\{(\nu_{23}+\nu_{24})+(\mu_{24}+\mu_{25}+\mu_{26})\}\nonumber \\
&\;\; - H(S_{2\alpha}^{n})\{({\mu}_{13}+\mu_{16}+\mu_{19})\}\;-\;
H(S_{2\beta}^{n})\{({\nu}_{12}+\nu_{14}+\nu_{16})\}\nonumber \\
&\;\; -
H(U_{2\alpha}^{n}|X_{2}^{n})\{({\nu}_{21}+{\nu}_{22})+({\mu}_{21}+{\mu}_{22}+{\mu}_{23})\}\nn\\
&\;\; -
H(U_{2\beta}^{n}|X_{2}^{n})\{(\nu_{23}+\nu_{24})+(\mu_{24}+\mu_{25}+\mu_{26})\}\nonumber \\
&\;\; -
H(S_{2\alpha}^{n}|U_{2\alpha}^{n})\{({\mu}_{11}+\mu_{14}+\mu_{17})\}\;-\;
H(S_{2\beta}^{n}|U_{2\beta}^{n})\{({\nu}_{11}+\nu_{13}+\nu_{15})\}\nonumber \\
&\;\; -
H(S_{2\alpha}^{n}|U_{2\beta}^{n})\{({\mu}_{12}+\mu_{15}+\mu_{18})\} \\
\nn \\
&\stackrel{\text{(b)}}{=} \sum_{i=1}^6
{\nu}_{1i}H(Y_{1\beta}^n|V_{1i\beta}^{n})+
\sum_{i=1}^9 {\mu}_{1i}H(Y_{1\alpha}^n|V_{1i\alpha}^{n}) \nonumber \\
&\;\; + H(U_{1\alpha}^{n})\{(\mu_{22}+\mu_{25}+\mu_{28})\}\;-\;
H(U_{1\beta}^{n})\{(\mu_{22}+\mu_{25}+\mu_{28})\}\nonumber \\
&\;\; -
H(S_{1\alpha}^{n}|U_{1\beta}^{n})\{(\mu_{22}+\mu_{25}+\mu_{28})\}\nonumber \\
&\;\; -
H(U_{1\alpha}^{n}|X_{1}^{n})\{(\mu_{22}+\mu_{25}+\mu_{28})+(\mu_{23}+\mu_{26}+\mu_{29})\}\nn\\
&\;\;-
H(U_{1\beta}^{n}|X_{1}^{n})\{(\nu_{22}+\nu_{24}+\nu_{26})-(\mu_{22}+\mu_{25}+\mu_{28})\}\nonumber \\
&\;\; -
H(S_{1\alpha}^{n}|U_{1\alpha}^{n})\{(\mu_{21}+\mu_{24}+\mu_{27})\}\;-\;
H(S_{1\beta}^{n}|U_{1\beta}^{n})\{(\nu_{21}+\nu_{23}+\nu_{25})\} \nonumber \\
&\;\; + \sum_{i=1}^6 {\nu}_{2i}H(Y_{2\beta}^n|V_{2i\beta}^{n})+
\sum_{i=1}^9 {\mu}_{2i}H(Y_{2\alpha}^n|V_{2i\alpha}^{n}) \nonumber \\
&\;\; + H(U_{2\alpha}^{n})\{(\mu_{12}+\mu_{15}+\mu_{18})\}\;-\;
H(U_{2\beta}^{n})\{(\mu_{12}+\mu_{15}+\mu_{18})\}\nonumber \\
&\;\; -
H(S_{2\alpha}^{n}|U_{2\beta}^{n})\{(\mu_{12}+\mu_{15}+\mu_{18})\}\nonumber \\
&\;\; -
H(U_{2\alpha}^{n}|X_{2}^{n})\{(\mu_{12}+\mu_{15}+\mu_{18})+(\mu_{13}+\mu_{16}+\mu_{19})\}\nn\\
&\;\;-
H(U_{2\beta}^{n}|X_{2}^{n})\{(\nu_{12}+\nu_{14}+\nu_{16})-(\mu_{12}+\mu_{15}+\mu_{18})\}\nonumber \\
&\;\; -
H(S_{2\alpha}^{n}|U_{2\alpha}^{n})\{(\mu_{11}+\mu_{14}+\mu_{17})\}\;-\;
H(S_{2\beta}^{n}|U_{2\beta}^{n})\{(\nu_{11}+\nu_{13}+\nu_{15})\}
\end{align}
\begin{align}
&\stackrel{\text{(c)}}{\leq} \sum_{i=1}^6
{\nu}_{1i}H(Y_{1\beta}^n|V_{1i\beta}^{n})+
\sum_{i=1}^9 {\mu}_{1i}H(Y_{1\alpha}^n|V_{1i\alpha}^{n}) \nonumber \\
&\;\; -
H(U_{1\alpha}^{n}|X_{1}^{n})\{(\mu_{22}+\mu_{25}+\mu_{28})+(\mu_{23}+\mu_{26}+\mu_{29})\}\;-\;
H(U_{1\beta}^{n}|X_{1}^{n})\{(\nu_{22}+\nu_{24}+\nu_{26})\}\nonumber \\
&\;\; -
H(S_{1\alpha}^{n}|X_{1}^{n})\{(\mu_{21}+\mu_{24}+\mu_{27})\}\;-\;
H(S_{1\beta}^{n}|X_{1}^{n})\{(\nu_{21}+\nu_{23}+\nu_{25})\} \nonumber \\
&\;\; + \sum_{i=1}^6 {\nu}_{2i}H(Y_{2\beta}^n|V_{2i\beta}^{n})+
\sum_{i=1}^9 {\mu}_{2i}H(Y_{2\alpha}^n|V_{2i\alpha}^{n}) \nonumber \\
&\;\; -
H(U_{2\alpha}^{n}|X_{2}^{n})\{(\mu_{12}+\mu_{15}+\mu_{18})+(\mu_{13}+\mu_{16}+\mu_{19})\}\;-\;
H(U_{2\beta}^{n}|X_{2}^{n})\{(\nu_{12}+\nu_{14}+\nu_{16})\}\nonumber \\
&\;\; -
H(S_{2\alpha}^{n}|X_{2}^{n})\{(\mu_{11}+\mu_{14}+\mu_{17})\}\;-\;
H(S_{2\beta}^{n}|X_{2}^{n})\{(\nu_{11}+\nu_{13}+\nu_{15})\} \\
\nn \\
&\stackrel{\text{(d)}}{=} \sum_{i=1}^6
\nu_{1i}H(Y_{1\beta}^n|V_{1i\beta}^{n})+
\sum_{i=1}^9 \mu_{1i}H(Y_{1\alpha}^n|V_{1i\alpha}^{n}) \nonumber \\
&\;\; - H(S_{1\alpha}^{n}|X_{1}^{n})\lp \sum_{i=1}^9 \mu_{2i}
\rp\;-\;
H(S_{1\beta}^{n}|X_{1}^{n})\lp \sum_{i=1}^6 \nu_{2i} \rp\nonumber \\
&\;\; + \sum_{i=1}^6 \nu_{2i}H(Y_{2\beta}^n|V_{2i\beta}^{n})+
\sum_{i=1}^9 \mu_{2i}H(Y_{2\alpha}^n|V_{2i\alpha}^{n}) \nonumber \\
&\;\; - H(S_{2\alpha}^{n}|X_{2}^{n})\lp \sum_{i=1}^9 \mu_{1i} \rp
\;-\; H(S_{2\beta}^{n}|X_{2}^{n})\lp \sum_{i=1}^6 \nu_{1i} \rp.
\end{align}
Here,
\begin{itemize}
\item to get inequality (a), we used \eqref{obeq1b1}-\eqref{obeq2a9}
in \eqref{eq:upperbound} and collected the terms together;
\item for equality (b), we used Proposition~\ref{prop:dualsol} along
with the facts
\beqa
H(U_{1\alpha}^n)&=&H(S_{1\alpha}^n),\\
H(U_{1\beta}^n)&=&H(S_{1\beta}^n),\\
H(U_{2\alpha}^n)&=&H(S_{2\alpha}^n)\\
H(U_{2\beta}^n)&=&H(S_{2\beta}^n);
\eeqa
\item inequality  (c) follows from the fact that
conditioning reduces entropy. In particular,
\begin{align}
H(U_{1\alpha}^n)-H(U_{1\beta}^n)-H(S_{1\alpha}^n|U_{1\beta}^n) &= -
H(U_{1\beta}^n|S_{1\alpha}^n) &\leq - H(U_{1\beta}^n|X_1^n) \\
H(U_{2\alpha}^n)-H(U_{2\beta}^n)-H(S_{2\alpha}^n|U_{2\beta}^n) &= -
H(U_{2\beta}^n|S_{2\alpha}^n) &\leq - H(U_{2\beta}^n|X_2^n);
\end{align}
\item for
equality (d) we used
\begin{align*}
H(U_{1\alpha}^n|X_1^n)=H(S_{1\alpha}^n|X_1^n),\quad&
H(U_{1\beta}^n|X_1^n)=H(S_{1\beta}^n|X_1^n)\\
H(U_{2\alpha}^n|X_2^n)=H(S_{2\alpha}^n|X_2^n),\quad&
H(U_{2\beta}^n|X_2^n)=H(S_{2\beta}^n|X_2^n).
\end{align*}
\end{itemize}
Now we single-letterize using the chain rule along with the fact
that the channel is memoryless and conditioning reduces entropy.
\begin{align}
aR_{1}+bR_{2}-(a+b)\epsilon_{n} &{\leq} \frac{1}{n} \sum_{q=1}^{n}
\left\{ \sum_{i=1}^6 \nu_{1i}H(Y_{1\beta}(q)|V_{1i\beta}(q))+
\sum_{i=1}^9 \mu_{1i}H(Y_{1\alpha}(q)|V_{1i\alpha}(q)) \nonumber \right.\\
&\;\; -
H(S_{1\alpha}(q)|X_{1}(q))\lp \sum_{i=1}^9 \mu_{2i} \rp\nn\\
&\;\; -
H(S_{1\beta}(q)|X_{1}(q))\lp \sum_{i=1}^6 \nu_{2i} \rp\nonumber \\
&\;\; + \sum_{i=1}^6 \nu_{2i}H(Y_{2\beta}(q)|V_{2i\beta}(q))+
\sum_{i=1}^9 \mu_{2i}H(Y_{2\alpha}(q)|V_{2i\alpha}(q)) \nonumber \\
&\;\; -
H(S_{2\alpha}(q)|X_{2}(q))\lp \sum_{i=1}^9 \mu_{1i} \rp \nn\\
&\;\; - \left. H(S_{2\beta}(q)|X_{2}(q))\lp \sum_{i=1}^6 \nu_{1i}
\rp \right\}.
\end{align}
\begin{align}
aR_{1}+bR_{2}-(a+b)\epsilon_{n} &\leq \sum_{i=1}^6
\nu_{1i}H(Y_{1\beta}|V_{1i\beta},Q)+
\sum_{i=1}^9 \mu_{1i}H(Y_{1\alpha}|V_{1i\alpha},Q) \nonumber \\
&\;\;\; - \left(\sum_{i=1}^9 \mu_{2i}\right) H(S_{1\alpha}|X_{1},Q)
-
\left(\sum_{i=1}^6 \nu_{2i}\right) H(S_{1\beta}|X_{1},Q)\nonumber \\
&\;\;\; + \sum_{i=1}^6 \nu_{2i}H(Y_{2\beta}|V_{2i\beta},Q)+
\sum_{i=1}^9 \mu_{2i}H(Y_{2\alpha}|V_{2i\alpha},Q) \nonumber \\
&\;\;\; - \left(\sum_{i=1}^9 \mu_{1i}\right) H(S_{2\alpha}|X_{2},Q)
-
\left(\sum_{i=1}^6 \nu_{1i}\right) H(S_{2\beta}|X_{2},Q) \nonumber \\
\nonumber \\
&\df c^{(out)}_{(\la,a,b)}(Q,X_{1},X_{2}),\label{eq:cout}
\end{align}
where we set $\lp Q, U_{1\beta}, U_{1\alpha}, X_1, S_{1\beta},
S_{1\alpha}, U_{2\beta}, U_{2\alpha}, X_2, S_{2\beta}, S_{2\alpha}
\rp$ to be joint random variables such that $Q$ is uniformly
distributed over $\{1,2,\ldots,n\}$ and,
\begin{align}
&\text{Pr}\lp U_{1\beta},U_{1\alpha},X_1,S_{1\alpha},S_{1\beta},U_{2\beta},U_{2\alpha},X_2,S_{2\alpha},S_{2\beta} |Q=q \rp\nn\\
&\;=\text{Pr}\lp
U_{1\beta}(q),U_{1\alpha}(q),X_1(q),S_{1\alpha}(q),S_{1\beta}(q),U_{2\beta}(q),U_{2\alpha}(q),X_2(q),S_{2\alpha}(q),S_{2\beta}(q)
\rp,
\end{align}
for $1\leq q\leq n$. Since the messages are independent for the two
users, so are $X_{1}(q)$ and $X_{2}(q)$. Therefore,
$(Q,X_{1},X_{2})$ satisfies the Markov chain $X_{1}-Q-X_{2}$.
Further because of our choice of $(U_{1\beta}(q),U_{1\alpha}(q),
U_{1\beta}(q),U_{1\alpha}(q))$, the random variables satisfy the
condition~\eqref{eq:aux}. Hence the random variables
$({Q},{X}_1,U_{1\alpha},U_{1\beta},{X}_2,U_{2\alpha},U_{2\beta})$
belong to the sub-family of ${\mathcal P}$ that we described
earlier, whose elements are defined by $(Q,X_{1},X_{2})$.

We are now ready to formally define $\mathcal R_{out}({Q},{X}_{1},{X}_{2})$:

\begin{align}
\mathcal R_{out}({Q},{X}_{1},{X}_{2})
\df \{(R_{1},R_{2})&:\;R_{1}\geq0,\;R_{2}\geq0 \nn\\
&aR_{1}+bR_{2} \leq c^{(out)}_{(\la,a,b)}({Q},{X}_{1},{X}_{2}),
\;\forall \la \in \La_{(a,b)}^{\prime},\;\forall a\geq0,\;b\geq0\}.
\nn
\end{align}

We have proved that if $(R_{1},R_{2})$ is achievable, then
\beq
(R_{1},R_{2}) \in  \bigcup_{{Q},{X}_{1},{X}_{2}} {\mathcal R}_{out}({Q},{X}_{1},{X}_{2}).
\eeq
This completes the proof.
\hfill$\Box$

\subsection{Proof Of Theorem~\ref{thm:Main2} (ii)}\label{sec:thm2-2}

For a given $\lp Q, X_{1}, X_{2} \rp$  such that ${X}_1 - {Q} -
{X}_2$ is a Markov chain, we need to quantify the gap between
${\mathcal R}_{out}({Q},{X}_{1},{X}_{2})$ and ${\mathcal
R}_{in}({Q},{X}_{1},{X}_{2})$, which are defined by Equations~\eqref{eq:Rin}
and~\eqref{eq:Rout} respectively. In order to do this, we quantify
the gap between $c^{(out)}_{(\la,a,b)}({Q},{X}_{1},{X}_{2})$ and
$c^{(in)}_{(\la,a,b)}({Q},{X}_{1},{X}_{2})$.

\begin{align}
&c^{(out)}_{(\la,a,b)}({Q},{X}_{1},{X}_{2})\;-\;c^{(in)}_{(\la,a,b)}({Q},{X}_{1},{X}_{2})\nn\\
&\;=(H(S_{1\alpha}|U_{1\alpha},Q)-H(S_{1\alpha}|X_{1},Q))
 \left(\sum_{i=1}^9 \mu_{2i}\right)
+ (H(S_{1\beta}|U_{1\beta},Q)-H(S_{1\beta}|X_{1},Q))
 \left(\sum_{i=1}^6 \nu_{2i}\right) \nonumber \\
&\;\;\;+ (H(S_{2\alpha}|U_{2\alpha},Q)-H(S_{2\alpha}|X_{2},Q))
 \left(\sum_{i=1}^9 \mu_{1i}\right)
 +
(H(S_{2\beta}|U_{2\beta},Q)-H(S_{2\beta}|X_{2},Q))
 \left(\sum_{i=1}^6 \nu_{1i}\right)\nonumber \\
&\;= I(S_{1\alpha};X_{1}|U_{1\alpha},Q)
 \left(\sum_{i=1}^9 \mu_{2i}\right) +
I(S_{1\beta};X_{1}|U_{1\beta},Q)
 \left(\sum_{i=1}^6 \nu_{2i}\right) \nonumber \\
&\quad + I(S_{2\alpha};X_{2}|U_{2\alpha},Q)
 \left(\sum_{i=1}^9 \mu_{1i}\right) +
I(S_{2\beta};X_{2}|U_{2\beta},Q)
 \left(\sum_{i=1}^6 \nu_{1i}\right)\nonumber\\
&\;\leq
b \max(I(S_{1\alpha};X_{1}|U_{1\alpha},Q),I(S_{1\beta};X_{1}|U_{1\beta},Q)) \nonumber\\
&\quad +
a \max(I(S_{2\alpha};X_{2}|U_{2\alpha},Q),I(S_{2\beta};X_{2}|U_{2\beta},Q)) \nonumber \\
&\;\leq
 a\Delta_1(Q,X_1,X_2)+b\Delta_2(Q,X_1,X_2).
\end{align}
Here $\Delta_1(Q,X_1,X_2)$ and  $\Delta_2(Q,X_1,X_2)$ are defined
as follows:
\begin{align*}
\Delta_1(Q,X_1,X_2)&\df\max(I(S_{2\alpha};X_{2}|U_{2\alpha}),
   I(S_{2\beta};X_{2}|U_{2\beta})),\\
\Delta_2(Q,X_1,X_2)&\df\max(I(S_{1\alpha};X_{1}|U_{1\alpha}),
   I(S_{1\beta};X_{1}|U_{1\beta})).
\end{align*}
This completes the  proof of  Theorem~\ref{thm:Main2}. \hfill$\Box$

\subsection{Proof Of Corollary \ref{cor:gcodebook}} \label{sec:proofgcodebook}

Consider the  $2$-state compound Gaussian
interference channel. For this special case, we have the following
result that identifies the Gaussian code books to be sufficient.
\begin{lem}\label{lemma:outerboundlem}
\beq {\mathcal R}_{out}({Q},{X}_{1},{X}_{2}) \subseteq {\mathcal
R}_{out}({Q}^{*},{X}_{1}^{*},{X}_{2}^{*}) \eeq where
$Q^{*}=1,\;X_{1}^{*}\sim{\mathcal
CN}(0,P_{1}),\;X_{2}^{*}\sim{\mathcal CN}(0,P_{2})$.
\end{lem}
We note for easy reference that ${\mathcal R}_{out}({Q},{X}_{1},{X}_{2})$ is defined
in Equation~\eqref{eq:Rout}.

\noindent\emph{Proof}:  It suffices to show that
\[ c^{(out)}_{(\la,a,b)}({Q},{X}_{1},{X}_{2}) \leq c^{(out)}_{(\la,a,b)}({Q}^{*},{X}_{1}^{*},{X}_{2}^{*}),\]
where $c^{(out)}_{(\la,a,b)}({Q},{X}_{1},{X}_{2})$ is as defined in
\eqref{eq:cout}.
\begin{align}
c^{(out)}_{(\la,a,b)}({Q},{X}_{1},{X}_{2}) &= \sum_{i=1}^6
\nu_{1i}h(Y_{1\beta}|V_{1i\beta},Q)+
\sum_{i=1}^9 \mu_{1i}h(Y_{1\alpha}|V_{1i\alpha},Q) \nonumber \\
&\;\;\; - \left(\sum_{i=1}^9 \mu_{2i}\right) h(S_{1\alpha}|X_{1},Q)
-
\left(\sum_{i=1}^6 \nu_{2i}\right) h(S_{1\beta}|X_{1},Q)\nonumber \\
&\;\;\; + \sum_{i=1}^6 \nu_{2i}h(Y_{2\beta}|V_{2i\beta},Q)+
\sum_{i=1}^9 \mu_{2i}h(Y_{2\alpha}|V_{2i\alpha},Q) \nonumber \\
&\;\;\; - \left(\sum_{i=1}^9 \mu_{1i}\right) h(S_{2\alpha}|X_{2},Q)
- \left(\sum_{i=1}^6 \nu_{1i}\right) h(S_{2\beta}|X_{2},Q) \nonumber
\end{align}
The terms $h(S_{1\alpha}|X_{1},Q), h(S_{1\beta}|X_{1},Q),
h(S_{2\alpha}|X_{2},Q)$ and $h(S_{2\beta}|X_{2},Q)$ are the
differential entropies of complex Gaussian noise with known variance and
are readily handled.  Let us now turn to the term $h(Y_{1\beta}|V_{11\beta},Q)$:
\begin{align}
h(Y_{1\beta}|V_{11\beta},Q) & = h(Y_{1\beta}|U_{1\alpha}U_{2\beta},Q) \\
& = \sum_{q} p(q) h(Y_{1\beta}|U_{1\alpha}U_{2\beta},Q=q) \\
& \stackrel{(a)}{\leq} \sum_{q} p(q) \log \lp \frac{P_{1q}}{|h_{12\alpha}|^{2}P_{1q}+1} + 1+\frac{|h_{21\beta}|^{2}P_{2q}}{|h_{21\beta}|^{2}P_{2q}+1} \rp \\
& \stackrel{(b)}{\leq} \log \lp \frac{P_{1}}{|h_{12\alpha}|^{2}P_{1}+1} + 1+\frac{|h_{21\beta}|^{2}P_{2}}{|h_{21\beta}|^{2}P_{2}+1} \rp \\
& = h(Y_{1\beta}^{*}|U_{1\alpha}^{*}U_{2\beta}^{*},Q^{*}).
\end{align}
Here,
\begin{itemize}
\item in step (a), we denoted
\beq
\E\Lbr|X_{k}|^{2}|Q=q\Rbr=P_{kq},\quad k=1,2
\eeq
and used the fact that conditional differential entropy is
maximized with the Gaussian distribution under a covariance constraint (Lemma~1 \cite{Tho87});
\item in step(b), we used Jensen's inequality.
\end{itemize}
A similar argument follows for the other terms. To conclude, we have shown
that
\beq  c^{(out)}_{(\la,a,b)}({Q},{X}_{1},{X}_{2})
 \leq c^{(out)}_{(\la,a,b)}({Q}^{*},{X}_{1}^{*},{X}_{2}^{*}).\eeq
This completes the proof.
\hfill $\Box$

Finally, we can readily see the proof of Corollary \ref{cor:gcodebook}. This is because,
\beq \bigcup_{(Q,X_{1},X_{2})} {\mathcal R}_{out}({Q},{X}_{1},{X}_{2})
= {\mathcal R}_{out}({Q}^{*},{X}_{1}^{*},{X}_{2}^{*}),\eeq
as a direct consequence of Lemma~\ref{lemma:outerboundlem}.

\section{Discussion: Insights On The Non-Compound Interference Channel}
\label{sec:discuss}

In this section we consider the non-compound interference channel model
introduced in \cite{TT07}; this is a
specific instance of our model and is obtained by setting
$\alpha=\beta$. Our results, when specialized to this instance provide
an alternative derivation of the  results of Chong et
al.~\cite{CMG06} and Tse and Telatar~\cite{TT07}. Below we briefly
sketch our results with an aim to
compare and contrast the different proofs.  The goal is not only
to give better insight into existing results, but also to give an idea
on how our new proof technique scales more naturally to the 2-state compound
interference channel (and in general to the $n$-state compound
interference channel that we will describe in the next section). We
first describe the achievable scheme and the inner bound.
Following that, we will describe the outer-bound, focusing on contrasts
between the different approaches.

\subsection{Achievable Scheme}
The special case of the noncompound version is obtained by setting\beq
S_{k\beta}=S_{k\alpha}=S_{k}
\eeq and, correspondingly,
\beq
U_{k\beta}=U_{k\alpha}=U_{k}
\eeq
 for $k=1,2$. We also set
\beq
R_{k\alpha}=0.
\eeq
 We rename $R_{k\beta}$ as $T_{k}$ and $R_{kp}$
as $S_{k}$ to be consistent with the notation of Chong et al.\ \cite{CMG06}.

The superposition achievable scheme can now be described by joint random variables
\beq
P=(Q,U_{1},X_{1},U_{2},X_{2})\eeq with the joint distribution factoring as
\beq
p(q)p(x_{1}|q)p(x_{2}|q)p(u_{1}|x_{1}q)p(u_{2}|x_{2}q).\eeq
From Section~\ref{sec:thm1}, it follows that any rate vector
$(S_{1},T_{1},S_{2},T_{2})$ that satisfies,
 \begin{align}
 S_{1} &\leq I(Y_{1};X_{1}|U_{1},U_{2},Q), \quad \text{if } S_{1}>0, \label{eq:11}\\
 T_{2} + S_{1} &\leq I(Y_{1};X_{1}U_{2}|U_{1},Q), \quad \text{if } S_{1}>0, \\
 T_{1} + S_{1} &\leq I(Y_{1};X_{1}|U_{2},Q), \quad \text{if } T_{1}+S_{1}>0, \\
 T_{2}+T_{1} + S_{1} &\leq I(Y_{1};X_{1}U_{2}|Q), \quad \text{if } T_{1}+S_{1}>0,  \\
 \nn\\
 S_{2} &\leq I(Y_{2};X_{2}|U_{2},U_{1},Q), \quad \text{if } S_{2}>0, \\
 T_{1} + S_{2} &\leq I(Y_{2};X_{2},U_{1}|U_{2},Q), \quad \text{if } S_{2}>0, \\
 T_{2} + S_{2} &\leq I(Y_{2};X_{2}|U_{1},Q), \quad \text{if } T_{2}+S_{2}>0, \\
 T_{1}+T_{2} + S_{2} &\leq I(Y_{2};X_{2},U_{1}|Q), \quad \text{if } T_{2}+S_{2}>0, \label{eq:24}\\
 \nn\\
 S_{1},T_{1},S_{2},T_{2} &\geq 0, \label{eq:nn}
 \end{align}
is achievable. Define the 4-dimensional region  \beq
\tilde{{\mathcal
R}}_{in}^{(4)}(P)\df  \lbr (S_{1},T_{1},S_{2},T_{2}):
\text{satisfies~\eqref{eq:11}-\eqref{eq:nn}} \rbr \eeq and its
projection on the 2-dimensional space
\beq
\tilde{{\mathcal R}}_{in}(P) \df \lbr (R_{1},R_{2}): R_k =
S_{k}+T_{k}, k=1,2\rbr.
\eeq

On the other hand,
define \beq {\mathcal R}_{in}^{(4)}(P) \df \lbr
(S_{1},T_{1},S_{2},T_{2}):
\text{satisfies~\eqref{eq:ibeq1}-\eqref{eq:ibeqnn}} \rbr. \eeq
\begin{align}
 S_{1} &\leq I(Y_{1};X_{1}|U_{1},U_{2},Q) \label{eq:ibeq1}\\
 T_{2} + S_{1} &\leq I(Y_{1};X_{1}U_{2}|U_{1},Q) \\
 T_{1} + S_{1} &\leq I(Y_{1};X_{1}|U_{2},Q) \\
 T_{2}+T_{1} + S_{1} &\leq I(Y_{1};X_{1}U_{2}|Q) \\
 \nn\\
 S_{2} &\leq I(Y_{2};X_{2}|U_{2},U_{1},Q) \\
 T_{1} + S_{2} &\leq I(Y_{2};X_{2},U_{1}|U_{2},Q) \\
 T_{2} + S_{2} &\leq I(Y_{2};X_{2}|U_{1},Q) \\
 T_{1}+T_{2} + S_{2} &\leq I(Y_{2};X_{2},U_{1}|Q) \label{eq:ibeq8}\\
 \nn\\
 S_{1}+T_{1} &\geq 0, \\
 S_{2}+T_{2} &\geq 0. \label{eq:ibeqnn}
 \end{align}
Let its projection on the 2-dimensional space
\beq
{\mathcal R}_{in}(P) \df \lbr (R_{1},R_{2}): R_k =
S_{k}+T_{k}, k=1,2\rbr.
\eeq
In Theorem~2 of \cite{CMG06}, the authors explicitly evaluated the constraints
that define this set and described it as the ``compact version" of the Han-Kobayashi
region \cite{HK81} (which results from a somewhat different coding strategy, as compared
to the superposition coding one). However, we know from
Lemma~\ref{lem:ib} that \beq
{\mathcal
R}_{in}(P)\subseteq\tilde{{\mathcal R}}_{in}(P).
\eeq
We thus conclude the alternate proof of ~\cite{CMG06}(Theorem 2). Our
 approach differs from the approach of Chong et
al.~\cite{CMG06} in two ways:
 \begin{itemize}
 \item It is instructive to observe
the similarities and differences between the 4-dimensional achievable region $\tilde{{\mathcal R}}_{in}^{(4)}(P)$ to the one in~\cite{CMG06}(Lemma 3). First, the inequalities involved are the same. Howevber,
 several of these constraints are inactive when the boundary conditions on the data rates bite.
We can immediately conclude that our achievable region ($\tilde{{\mathcal R}}_{in}^{(4)}(P)$)
is in general a superset of the region in~\cite{CMG06}. This is somewhat
surprising since the encoding method in
both cases is superposition coding. The differences result due to our careful
consideration of the error
events in the decoding process.
\item  Chong et al.\ \cite{CMG06} described the 2-dimensional region explicitly by carrying out the
somewhat tedious algorithmic procedure of Fourier-Motzkin elimination. Further, they showed that a potentially bigger region (the compact description region) is achievable by time-sharing between two other schemes defined by $(Q,\emptyset,X_{1},U_{2},X_{2})$ and $(Q,U_{1},X_{1},\emptyset,X_{2})$. In our approach,
 we entirely avoid describing the 2-dimensional region explicitly. Further, we showed that there is
 no need to time-share  between any other schemes, to achieve ${\mathcal R}_{in}(P)$.
 \end{itemize}

\subsection{Outer Bound}

For a given $P=(Q,U_{1},X_{1},U_{2},X_{2})$, the inner-bound region
in  Chong et al.\ \cite{CMG06} is described by seven linear
inequalities involving $R_{1}$ and $R_{2}$. In \cite{TT07}, Telatar and Tse  picked a
specific choice of $(U_{1},U_{2})$ given by
\beq
p(u_{1},u_{2}|q,x_{1},x_{2}) = p_{S_{1}|X_{1}}(u_{1}|x_{1})p_{S_{2}|X_{2}}(u_{2}|x_{2}).\eeq
In deriving the outer bound, Telatar and Tse~\cite{TT07} gave
extra information to the receivers (the so-called ``genie-aided" approach) to handle
the seven inequalities. The rationale to what  side information the genie
should provide to handle the different  linear inequalities was somewhat
speculative (cf.\ Section IV \cite{TT07}).

Our approach avoids an explicit representation of the
inner-bound. This higher level description allowed us (cf.\ Section~\ref{sec:extRep})
to show that any
inequality involved in the projected region can be obtained by linear combination of the
inequalities~\eqref{eq:ibeq1}-\eqref{eq:ibeq8}. Further, each
inequality in~\eqref{eq:ibeq1}-\eqref{eq:ibeq8} arises from a {\em typical
error event} consideration.  We now have the operational insight into what  side
information to give when. We demonstrate this process in the instance of
Equation~\eqref{eq:ibeq1}. This
inequality must be satisfied to ensure that the Receiver 1 decodes
its own private message,  on the condition that it can decode both
the public messages correctly. This suggests that corresponding to
this inequality, we may give the side information
$(U_{1}^{n},U_{2}^{n})$. A similar argument handles each of the other
inequalities~\eqref{eq:ibeq1}-\eqref{eq:ibeq8}.

\section{$N$-state Compound Interference Channel}\label{sec:gen}

In this section we consider the natural extension of the $2$-state compound
interference channel to an $N$-state
compound interference channel. Our earlier results (both inner and outer bounds)
 also generalize naturally  to the more general $N$-state model.

\subsection{Model}

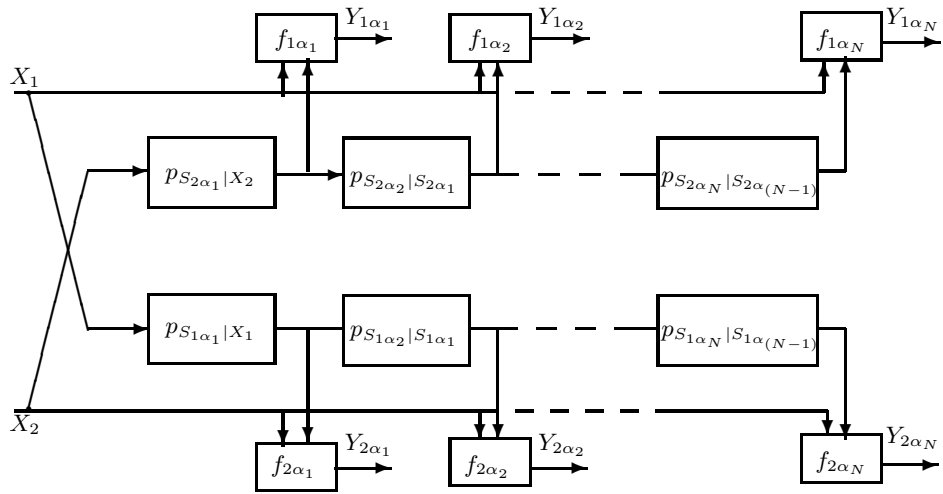
\begin{figure}[h]
\begin{center}
\scalebox{1.0}{\input{nCompoundIC.latex}}
\end{center}
\caption{The $N$-state compound interference channel model.}
\label{fig-ncICmodel}
\end{figure}

The $N$-state
compound interference channel is depicted in Figure~\ref{fig-ncICmodel}.
Each receiver can be in one of the $N$ possible states denoted by
$\alpha_{1},\alpha_{2},\ldots,\alpha_{N}$.

\subsection{Results}

We can characterize the inner bound and
outer bounds to the capacity region in a way similar to the $2$-state compound channel.

\subsubsection*{Inner Bound}
Our coding scheme is  $N+1$-level
superposition coding. This is much along the lines of the 3-level superposition coding
employed for the $2$-state compound interference
channel. The coding scheme is characterized by jointly distributed random variables
\beq
\lp Q,X_{1}, U_{1\alpha_{1}},\ldots,U_{1\alpha_{N}},X_{2}, U_{2\alpha_{1}},\ldots,U_{2\alpha_{N}}\rp
\eeq
which satisfy the Markov chain
\beq
U_{1\alpha_{N}}-\ldots-U_{1\alpha_{1}}-X_{1}-Q-X_{2}-U_{2\alpha_{1}}-\ldots-U_{2\alpha_{N}}.
\eeq
As earlier, we restrict ourselves to a subfamily of the jointly distributed random variables
uniquely determined by $(Q,X_{1},X_{2})$ in the following way:
\begin{quote}
Given $(Q,X_{1},X_{2})$, we pick random variables
\beq
\lbr U_{k\alpha_{n}},\;n=1,\ldots,N,\;k=1,2\rbr,
\eeq
 such that they have the same joint distribution as
\beq
\lbr S_{k\alpha_{n}},\;n=1,\ldots,N,\;k=1,2\rbr,
\eeq
but are independent of them.
\end{quote}
Using these random variables, we generate the $(N+1)$-level superposition
random code books for each user with rates
$(R_{1\alpha_{N}},\ldots,R_{1\alpha_{1}},R_{1p})$ and
$(R_{2\alpha_{N}},\ldots,R_{2\alpha_{1}},R_{2p})$ respectively.

The decoding at each receiver is jointly typical set decoding. It is
similar to the decoding described for the $2$-state.
Each receiver
tries to decode fully all of its own messages, but only partially
decodes the other (interfering) user. This strategy can be seen as
an opportunistic strategy where the extent of the interference that
the receiver decodes depends upon the level of interference it sees.

The remainder description of the achievable rate region follows the same
development pattern as for the 2-state compound channel. It would be
impractical (in terms of the length of the descriptions) to explicitly
detail this description. As such, we briefly itemize
the main points in the achievable region description below.
\begin{itemize}
\item We first have  an achievable rate region $\tilde{{\mathcal
R}}_{in}^{2(N+1)}(Q,X_{1},X_{2})$ in $2(N+1)$ dimensions along the same
lines as~\eqref{eq:tRin6} (we have avoided the explicit description of the
linear inequalities describing the region due to the tedium and length involved \
in doing so).  As earlier, let
$\tilde{{\mathcal R}}_{in}(Q,X_{1},X_{2})$ be the projection onto the 2-dimensional
space $(R_{1},R_{2})$ where,
\beq
R_{1}=R_{1\alpha_{N}}+\ldots+R_{1\alpha_{1}}+R_{1p},\quad R_{2}=R_{2\alpha_{N}}+\ldots+R_{2\alpha_{1}}+R_{2p}.
\eeq
We have that $\tilde{{\mathcal R}}_{in}(Q,X_{1},X_{2})$ is
achievable.
\item We next define ${\mathcal
R}_{in}^{2(N+1)}(Q,X_{1},X_{2})$ as a generalization of
\eqref{eq:Rin6} and define its projection onto the two dimensional
space ${\mathcal R}_{in}(Q,X_{1},X_{2})$. Lemma~\ref{lem:ib} can be appropriately
generalized to show that \beq {\mathcal
R}_{in}(Q,X_{1},X_{2})\subseteq\tilde{{\mathcal
R}}_{in}(Q,X_{1},X_{2}),\eeq
thus proving that ${\mathcal R}_{in}(Q,X_{1},X_{2})$ is also achievable.
\item We next characterize an extremal
representation of  ${\mathcal R}_{in}(Q,X_{1},X_{2})$, using an appropriate
generalization of~\eqref{eq:Rin} to the $N$-state model). In other
words, we represent it as an intersection of hyperplanes, where the
inequality used to define the hyperplane can be obtained as a
linear combination of the inequalities used to define ${\mathcal
R}_{in}^{2(N+1)}(Q,X_{1},X_{2})$.
\end{itemize}

\subsubsection*{Outer Bound}
An outer-bound
${\mathcal R}_{out}(Q,X_{1},X_{2})$ can be derived with an extremal representation
that is similar to the corresponding one for the inner bound
${\mathcal R}_{in}(Q,X_{1},X_{2})$ (this step is a natural generalization of
\eqref{eq:Rout}). In  deriving the outer bound, we
use appropriate  genie-aided techniques (that involve providing
 suitable side information to the receiver). Again, what
side information is shared is decided based on the
typical error events which lead to the corresponding inequality in the
inner bound.

\subsubsection*{Gap}
Finally, we characterize the gap between the outer and inner bounds
 to the capacity region for the $n$-state compound
channel, in much the same way as we did for the $2$-state compound
channel. This is stated formally below.
\begin{thm}\label{thm:nstate}
For the $N$-state compound interference channel of
Figure~\ref{fig-ncICmodel}, if $(R_1,R_2)$ is in the outer bound to
the capacity region, then $(R_1-\Delta_1,R_2-\Delta_2)$ is
achievable, where
\begin{align}
\Delta_1(Q,X_1,X_2)&=\max_{1\leq n \leq N}\quad I(X_2;S_{2\alpha_{n}}|U_{2\alpha_{n}}), \\
\Delta_2(Q,X_1,X_2)&=\max_{1\leq n \leq N}\quad
I(X_1;S_{1\alpha_{n}}|U_{1\alpha_{n}}).
\end{align}
\end{thm}

Specializing to the deterministic version, we can see that this gap
is zero and hence the capacity region is characterized exactly.
Specializing to the Gaussian version, we can see that this gap is no more than one bit.
This completes the extension to the $N$-state compound channel scenario.

\subsection{Discussion}
A few comments on the structure and properties of the achievable scheme are in order here.
\begin{itemize}
\item Note that the structure of the achievable scheme (or the power split in the
 Gaussian scheme), which is characterized by the joint random variables
\beq
\lp Q,X_{1}, U_{1\alpha_{1}},\ldots,U_{1\alpha_{N}},X_{2},
U_{2\alpha_{1}},\ldots,U_{2\alpha_{N}}\rp,
\eeq
depends {\em only on the
interference states and not on the deterministic functions}
$f_{k\alpha_{n}}$. The functions $f_{k\alpha_{n}}$ however may still help in
determining the actual achievable rate region.

We highlight this point by  considering the case when each of the degraded interference
channels in our model are identity, i.e.,
\beq S_{k\alpha_{2}}=S_{k\alpha_{1}}=\ldots=S_{k\alpha_{N}}, \quad k=1,2.\eeq
For this model the ``compoundness" of the channel is only due to the
functions $f_{k\alpha_{n}}$. Indeed,  only two levels of
superposition coding suffice, much as in the noncompound version of the problem.

\item Let us assume
\beq
S_{1\alpha_{2}}=S_{1\alpha_{1}}.
\eeq
Then  our
 achievable scheme sets \beq
 U_{1\alpha_{2}}=U_{1\alpha_{1}}.
 \eeq
  This implies that the level of the code book corresponding to $U_{1\alpha_{2}}$ is
 ``degenerate" and that we might as well set
 \beq
 R_{1\alpha_{2}}=0.
 \eeq
  Suppose, however that
 \beq
 f_{1\alpha_{2}}\not=f_{1\alpha_{1}}\eeq
  and hence the two receiver states
 $Y_{1\alpha_{1}}$ and $Y_{1\alpha_{2}}$ are not the same. While the receiver in
 either state adopts the same decoding technique  (with respect to the level of
 interference it decodes),
the higher dimensional constraints on the rate vector,
as imposed by the  decoding condition for  each state, are different. Nevertheless,
 we see that for the Gaussian case one of these states is always worse than the other
 and thus would be the critical bottleneck in determining the achievable rates; this is
 done next.
\end{itemize}

\section{The Compound Gaussian Interference Channel}\label{sec:Gaussian}

\subsection{Model}
The scalar Gaussian interference channel is defined by the complex
channel parameters $(h_{11},h_{21},h_{12},h_{22})$. The finite state
compound channel version of it allows these parameters to take
values in a finite set $\A$. \beq \A = \lbr
(h_{11},h_{21},h_{12},h_{22})_1, (h_{11},h_{21},h_{12},h_{22})_2, \ldots ,
 (h_{11},h_{21},h_{12},h_{22})_{|\A|}\rbr. \eeq
 Define \beq \A_{k}
\df \lbr (h_{1k},h_{2k})
|\;(h_{11},h_{21},h_{12},h_{22})\in\A \rbr,\quad k=1,2, \eeq
Observe that the channels from the two transmitters to the receiver $k$ are
defined solely by the parameters
$(h_{1k},h_{2k})$. Therefore the set $A_{k}$ is the set of states that
the receiver $k$ can take. Now define $\A^{\prime}$ as
\beqa
\A^{\prime} &\df& \A_{1}\times \A_{2} \nn\\
&=& \lbr (h_{11},h_{21},h_{12},h_{22}) |\;
(h_{11},h_{21})\in\A_{1},\;(h_{12},h_{22})\in\A_{2} \rbr. \eeqa
In other words $\A^{\prime}$ allows for {\em all combinations} of the possible
states for both the receivers. Let ${\mathcal C}(\A)$ denote
the capacity region of the compound channel defined by the set $\A$.
We have the following proposition:

\begin{prop}
\[{\mathcal C}(\A) = {\mathcal C}(\A^{\prime}).\]
\end{prop}

\noindent{\em Proof:}
Note that $\A \subseteq \A^{\prime}$.
Thus it is clear that any scheme that works for the compound channel
$\A^{\prime}$ also works for the compound channel $\A$. However,
since the two receivers do not cooperate, only the {\em marginal} channels
to each receiver decides the decodability of any communication scheme. We
now conclude that a scheme that works for
the compound channel $\A$ also works for the compound channel
$\A^{\prime}$. This completes the proof.
\hfill$\Box$

In the light of this observation, we can, without loss of
generality, consider finite state compound channels whose state set
$\A$ decomposes as $\A_{1}\times\A_{2}$.

In Section~\ref{sec:model} we saw that the case where the
cardinality of $\A_{1}$ and $\A_{2}$ is restricted to $2$ is
captured by the $2$-state compound interference channel of
Figure~\ref{fig-cICmodel}. Analogously, the general case where
$|\A_{1}|$ and $|\A_{2}|$ are finite (with cardinality no more than $N$)
is captured by the $N$-state
compound interference channel of Figure~\ref{fig-ncICmodel}. We see this
formally below. The key point is the infinitely divisible nature of Gaussian
statistics. This aspect was used to show that
the scalar Gaussian broadcast channel is always stochastically
degraded. In a similar vein, the compound scalar Gaussian interference
channel can always be supposed to have degraded interference states.

We begin by noting that if \beq
|\A_{1}|\not=|\A_{2}|,
\eeq
then we can add
redundant duplicate copies to one of the sets, so that
\beq
|\A_{1}|=|\A_{2}|.
\eeq
Therefore, without of loss of generality, we
suppose this is true.:
\beq
|\A_{1}|=|\A_{2}|=N.
\eeq
Then, {\em order} the finite sets
$\A_{1}$ and $\A_{2}$ such that
\begin{align*}
|h_{21\alpha_{1}}|\geq |h_{21\alpha_{2}}|\geq\ldots\geq|h_{21\alpha_{N}}|, \\
|h_{12\alpha_{1}}|\geq
|h_{12\alpha_{2}}|\geq\ldots\geq|h_{12\alpha_{N}}|.
\end{align*}
Next, we do the following substitution to reduce the finite state
Gaussian interference channel to the model of
Figure~\ref{fig-ncICmodel}.
\begin{align}
S_{1\alpha_{1}} &= h_{12\alpha_{1}}X_{1}+Z_{1\alpha_{1}}, \\
S_{1\alpha_{n}} &=
\frac{h_{12\alpha_{n}}}{h_{12\alpha_{(n-1)}}}S_{1\alpha_{(n-1)}}+\lp 1 - \left| \frac{h_{12\alpha_{n}}}{h_{12\alpha_{(n-1)}}}\right|^{2} \rp^{1/2} Z_{1\alpha_{n}}, \quad n=2,\ldots,N,\\
S_{2\alpha_{1}} &= h_{21\alpha_{1}}X_{2}+Z_{2\alpha_{1}}, \\
S_{2\alpha_{n}} &=
\frac{h_{21\alpha_{n}}}{h_{21\alpha_{(n-1)}}}S_{2\alpha_{(n-1)}}+\lp
1 - \left| \frac{h_{21\alpha_{n}}}{h_{21\alpha_{(n-1)}}}\right|^{2}
\rp^{1/2} Z_{2\alpha_{n}}, \quad n = 2,\ldots,N.
\end{align}
\begin{align}
Y_{1\alpha_{n}} &=f_{1\alpha_{n}}(X_{1},S_{2\alpha_{n}})= h_{11\alpha_{n}}X_{1}+S_{2\alpha_{n}},\quad 1\leq n\leq N,\\
Y_{2\alpha_{n}} &=f_{2\alpha_{n}}(X_{2},S_{1\alpha_{n}})=
h_{22\alpha_{n}}X_{2}+S_{1\alpha_{n}},\quad 1\leq n\leq N.
\end{align}
Here $Z_{k\alpha_{n}}$s are independent complex Gaussian  random
variables with unit variance. Note that the function
$f_{k\alpha_{n}}$ captures the direct link gains $h_{11}$ and
$h_{22}$. The channels
$p(S_{k\alpha_{n}}|S_{k\alpha_{(n-1)}})$ capture the cross link
gains $h_{12}$ and $h_{21}$ as well as the additive noise.

\subsection{Main Result}
We are now ready to summarize the main result of this paper.
\begin{thm}
For the finite state compound Gaussian interference channel,  multilevel
superposition coding with Gaussian
code books and opportunistic decoding depending on the interference state
is within one bit of the capacity region.
\end{thm}
\noindent\emph{Proof}: We have shown earlier in this section that any
finite state Gaussian interference channel is captured as a special
case of the model in Figure \ref{fig-ncICmodel}. Specializing the
result of Theorem \ref{thm:nstate} to the Gaussian case, we have
that the multilevel superposition coding is within one bit to the
capacity. Further,  it suffices to only consider
 Gaussian code books in the superposition code (along the same lines as Corollary
\ref{cor:gcodebook}).
\hfill$\Box$

\subsection{Discussion}
A few remarks are in order now.
\begin{enumerate}
\item While we have restricted both the direct-link gains ($h_{11}$ and $h_{22}$) and
the cross-link gains ($h_{12}$ and $h_{21}$) to a finite set of values so far, it turns
out that we can relax this assumption for the direct link gain.
In particular, suppose
\beq
(h_{11},h_{21}), (h_{11}^{\prime},h_{21}) \in \A_{1},\quad |h_{11}|<|h_{11}^{\prime}|.
\eeq
These correspond to two states of the receiver $1$, which
differ only in the direct link gain, but have the same cross link
gain. As observed in the previous section, for either of the two states, the
receiver adopts the {\em same} decoding method. Further, since we have
restricted ourselves to Gaussian code books, we see that the performance is
restricted only by the  state that has the weaker of the two direct links:
in this  case, it is the one  with parameters $(h_{11},h_{21})$
To summarize:
\begin{quote}
If a rate vector is achievable for the state
$(h_{11},h_{21})$, then it is also achievable for
$(h_{11}^{\prime},h_{21})$.
\end{quote}
Therefore, at any receiver, for a fixed cross-link  value the
direct-link gain at the corresponding receiver
can take on values in a set. This set could be infinite. The receiver
state that has the weakest direct link is the bottleneck.

\item At any receiver, the set of unique values the cross-link gains
take is assumed to be finite. This
finiteness assumption appears to be critical since the number of levels
required in the superposition coding at the interfering transmitter
depends on it. If the cardinality of the set is $N$, then the number of
levels in the superposition coding is $N+1$. It is an open question to determine
if a finite number of levels of superposition coding suffice to handle a
continuum of interference cross-link gains.

\end{enumerate}

\appendix

\section{Analysis Of Probability Of Error} \label{app:Pe}

In the following we consider the decodability conditions at receiver
$\text{Rx}_{1\beta}$ only. A very similar analysis applies to the other
receiver-state pairs.

Due to the symmetry of the random code book generation, the probability of
error averaged over the ensemble of random random code books, does not depend on
which codeword was sent. Hence, without loss of generality, we can
assume that the messages indexed by
\beq
(j_{1},k_{1},l_{1})=(1,1,1),\quad (j_{2},k_{2},l_{2})=
(1,1,1),
\eeq
were sent by the two transmitters respectively. Let us define the following event
\begin{align*}
E_{jklm} &= \\
&\lbr \lp Q^{n}, U_{1\beta}^{n}\lp j\rp,U_{1\alpha}^{n}\lp
j,k\rp,X_{1}^{n}\lp j,k,l\rp,U_{2\beta}^{n}\lp
m\rp,Y_{1\beta}^{n}\rp \in A^{\lp n\rp}_{\epsilon} \lp Q,
U_{1\beta},U_{1\alpha},X_{1},U_{2\beta},Y_{1\beta}\rp \rbr.
\end{align*}
Letting $P^{(n)}_{e}$ denote the probability of decoding error at
$\text{Rx}_{1\beta}$ we have
\begin{align}
P^{(n)}_{e} &= P \lp \lp \cup_{m}E_{111m} \rp ^{c} \bigcup \cup_{(j,k,l)\not=(1,1,1)}E_{jklm} \rp \\
&{\leq} \underbrace{P\lp\lp \cup_{m}E_{111m} \rp ^{c}\rp}_{(a)} \nn\\
&\quad + \underbrace{\sum_{l\not=1} P(E_{11l1})}_{(b)}
 + \underbrace{\sum_{l\not=1,m\not=1} P(E_{11lm})}_{(c)} \nn \\
&\quad+ \underbrace{\sum_{k\not=1,l} P(E_{1kl1})}_{(d)} + \underbrace{\sum_{k\not=1,l,m\not=1} P(E_{1klm})}_{(e)} \nn \\
&\quad+ \underbrace{\sum_{j\not=1,k,l} P(E_{jkl1})}_{(f)} +
\underbrace{\sum_{j\not=1,k,l,m\not=1} P(E_{jklm})}_{(g)}.
\label{eq:errevents}
\end{align}
The final inequality used the union bound. Let us
consider each term in~\eqref{eq:errevents} and study the
conditions needed to make it go to 0 asymptotically (in $n$).
\begin{itemize}
\item It
is straightforward to see that $(a)$ goes to 0 as $n\rightarrow\infty$.
\item Now consider $(b)$. We begin by noting that $l\in\{1,\ldots,2^{nR_{1p}}\}$.
Therefore if $R_{1p}=0$ then $(b)=0$. Else,  \beq (b) \leq
2^{nR_{1p}}
2^{-n(I(Y_{1\beta};X_{1}|U_{1\alpha},U_{2\beta},Q)-5\epsilon)}.\eeq
Therefore for $(b)$ to go to $0$ as $n\rightarrow\infty$, we must
have \beq R_{1p}\leq
I(Y_{1\beta};X_{1}|U_{1\alpha},U_{2\beta},Q),\quad \text{if
}R_{1p}>0. \label{eq:nogoodname1} \eeq
\item Similarly, $(c)$ is $0$ if
$R_{1p}=0$ or  $R_{2\beta}=0$. Else, it must be that \beq
R_{2\beta}+R_{1p} \leq I(Y_{1\beta};X_{1},U_{2\beta}|U_{1\alpha},Q).
\label{eq:nogoodname2}\eeq It is important to note that if $R_{2\beta}=0$, but
$R_{1p}>0$ then, \eqref{eq:nogoodname2} is redundant because of
\eqref{eq:nogoodname1}. Therefore for $(c)$ to go to $0$ as
$n\rightarrow\infty$ (assuming that $(b)$ goes to $0$ too), we must
have, \beq R_{2\beta}+R_{1p} \leq
I(Y_{1\beta};X_{1},U_{2\beta}|U_{1\alpha},Q),\quad \text{if
}R_{1p}>0. \eeq
\end{itemize}

Similarly for $(d),(e),(f)$ and $(g)$, we must have
\begin{align}
R_{1\alpha}+R_{1p} &\leq
I(Y_{1\beta};X_{1}|U_{1\beta},U_{2\beta},Q),\quad \text{if }R_{1\alpha}+R_{1p}>0,    \\
R_{2\beta}+R_{1\alpha}+R_{1p} &\leq
I(Y_{1\beta};X_{1},U_{2\beta}|U_{1\beta},Q),\quad \text{if }R_{1\alpha}+R_{1p}>0, \\
R_{1\beta}+R_{1\alpha}+R_{1p} &\leq
I(Y_{1\beta};X_{1}|U_{2\beta},Q),\quad \text{if }R_{1\beta}+R_{1\alpha}+R_{1p}>0, \\
R_{2\beta}+R_{1\beta}+R_{1\alpha}+R_{1p} &\leq
I(Y_{1\beta};X_{1},U_{2\beta}|Q),\quad \text{if
}R_{1\beta}+R_{1\alpha}+R_{1p}>0,
\end{align}
respectively.

\section{Proof Of Lemma~\ref{lem:ib}} \label{app:Rp}

Consider any $(R_{1},R_{2})\in{\mathcal R}_{in}(P)$. Then there
exists an
\beq
(R_{1p},R_{1\alpha},R_{1\beta},R_{2p},R_{2\alpha},R_{2\beta}) \in
{\mathcal R}^{(6)}_{in}(P)
\eeq
 such that,
\beq
R_{1}=R_{1p}+R_{1\alpha}+R_{1\beta}\quad \text{and} \quad R_{2}=R_{2p}+R_{2\alpha}+R_{2\beta}.
\eeq
We will find a
\beq
(\tilde{R}_{1p},\tilde{R}_{1\alpha},\tilde{R}_{1\beta},\tilde{R}_{2p},\tilde{R}_{2\alpha},\tilde{R}_{2\beta})
\in \tilde{{\mathcal R}}^{(6)}_{in}(P)
\eeq
such that,
\begin{align*}
\tilde{R}_{1p}+\tilde{R}_{1\alpha}+\tilde{R}_{1\beta} & = R_{1p}+R_{1\alpha}+R_{1\beta}, \\
\tilde{R}_{2p}+\tilde{R}_{2\alpha}+\tilde{R}_{2\beta} & = R_{2p}+R_{2\alpha}+R_{2\beta}, \\
\end{align*}
by the following algorithmic procedure.

\noindent\textbf{Step 1a)} For $k=1,2$, if $R_{k\beta}<0$ then,
\beq
R_{k\beta}\leftarrow0,\quad R_{k\alpha}\leftarrow R_{k\alpha}+R_{k\beta}.\eeq

\noindent\textbf{Step 1b)} For $k=1,2$, if $R_{k\alpha}<0$ then,
\beq
R_{k\alpha}\leftarrow0,\quad R_{kp}\leftarrow R_{kp}+R_{k\alpha}.\eeq

\noindent\textbf{Step 2a)} For $k=1,2$, if $R_{kp}<0$ then,
\beq
R_{kp}\leftarrow0,\quad R_{k\alpha}\leftarrow R_{k\alpha}+R_{kp}.\eeq

\noindent\textbf{Step 2b)} For $k=1,2$, if $R_{k\alpha}<0$ then,
\beq
R_{k\alpha}\leftarrow0,\quad R_{k\beta}\leftarrow R_{k\beta}+R_{k\alpha}.\eeq

First up, we note that at each step we are ensuring that
$R_{1p}+R_{1\alpha}+R_{1\beta}$ and $R_{2p}+R_{2\alpha}+R_{2\beta}$
stay invariant. Next, note that if
$R_{1p},R_{1\alpha},R_{1\beta},R_{2p},R_{2\alpha}$ and $R_{2\beta}$
are all nonnegative to begin with, then it is easy to see that
\beq
(R_{1p},R_{1\alpha},R_{1\beta},R_{2p},R_{2\alpha},R_{2\beta}) \in
\tilde{{\mathcal R}}^{(6)}_{in}(P)
\eeq
 and hence
\beq
(R_{1},R_{2})\in\tilde{{\mathcal R}}_{in}(P).
\eeq

\begin{claim}
At the end of \emph{Step 1b}, the new
$(R_{1p},R_{1\alpha},R_{1\beta},R_{2p},R_{2\alpha},R_{2\beta})$
still remains in ${\mathcal R}^{(6)}_{in}(P)$ and satisfies
\beq
R_{1\alpha},R_{1\beta},R_{2\alpha},R_{2\beta}\geq0.\eeq
\end{claim}
\noindent\emph{Proof:} Consider Step 1a. Note that in this step we
are potentially increasing $R_{k\beta}$, but the rest of the
components either remain the same or decrease. Also note that in
this step, we are keeping $R_{k\beta}+R_{k\alpha}$ invariant.
Therefore, we only need to ensure that the inequalities
among~\eqref{ibeq1b1}-\eqref{ibeq2a9} that have $R_{k\beta}$, but
not $R_{k\alpha}$ are not violated. This can be verified to be true,
because of the polymatroidal nature of each block of the
inequalities in~\eqref{ibeq1b1}-\eqref{ibeq2a9}. The argument is
similar for Step 1b. \hfill$\Box$

\begin{claim}
At the end of \emph{Step 2b}, the new
$(R_{1p},R_{1\alpha},R_{1\beta},R_{2p},R_{2\alpha},R_{2\beta})$ is
in $\tilde{\mathcal R}^{(6)}_{in}(P)$.
\end{claim}
\noindent\emph{Proof:} Note that in Step 2a, the only component that
potentially increases is $R_{kp}$, and so we might be violating one
of the following
constraints:~\eqref{ibeq1b1},\eqref{ibeq1b2},\eqref{ibeq1a1}-\eqref{ibeq1a3},\eqref{ibeq2b1},\eqref{ibeq2b2}
and \eqref{ibeq2a1}-\eqref{ibeq2a3}. However, by setting $R_{kp}=0$,
these violated constraints no longer matter for  $\tilde{{\mathcal
R}}_{in}^{(6)}(P)$. The argument is similar for Step 2b. Note that
at the end of Step 2b, we have ensured that all the components are
nonnegative. \hfill$\Box$

\end{document}

%% file: GIC.latex
\setlength{\unitlength}{3947sp}%
\begingroup\makeatletter\ifx\SetFigFont\undefined%
\gdef\SetFigFont#1#2#3#4#5{%
  \reset@font\fontsize{#1}{#2pt}%
  \fontfamily{#3}\fontseries{#4}\fontshape{#5}%
  \selectfont}%
\fi\endgroup%
\begin{picture}(3630,3927)(136,-3682)
\put(2871,-2816){\makebox(0,0)[lb]{\smash{{\SetFigFont{12}{14.4}{\rmdefault}{\mddefault}{\updefault}{\color[rgb]{0,0,0}+}%
}}}}
{\color[rgb]{0,0,0}\thinlines
\put(2921,-2764){\circle{150}}
}%
{\color[rgb]{0,0,0}\put(901,-811){\vector( 1,-1){1950}}
}%
{\color[rgb]{0,0,0}\put(901,-2761){\vector( 1, 1){1950}}
}%
{\color[rgb]{0,0,0}\put(2926,-2911){\vector( 0, 1){ 75}}
\put(2926,-2911){\line( 0,-1){ 75}}
\put(2926,-2986){\line( 0,-1){ 75}}
\put(2926,-3061){\line( 0,-1){ 75}}
\put(2926,-3136){\line( 0,-1){ 75}}
\put(2926,-3211){\line( 0,-1){ 75}}
\put(2926,-3286){\line( 0,-1){ 75}}
\put(2926,-3361){\line( 0,-1){ 75}}
}%
{\color[rgb]{0,0,0}\put(2926,-61){\line( 0,-1){ 75}}
\put(2926,-136){\line( 0,-1){ 75}}
\put(2926,-211){\line( 0,-1){ 75}}
\put(2926,-286){\line( 0,-1){ 75}}
\put(2926,-361){\line( 0,-1){ 75}}
\put(2926,-436){\line( 0,-1){ 75}}
\put(2926,-511){\line( 0,-1){ 75}}
\put(2926,-586){\vector( 0,-1){ 75}}
}%
{\color[rgb]{0,0,0}\put(901,-811){\vector( 1, 0){1950}}
}%
{\color[rgb]{0,0,0}\put(901,-2761){\vector( 1, 0){1950}}
}%
{\color[rgb]{0,0,0}\put(3001,-2761){\vector( 1, 0){675}}
}%
{\color[rgb]{0,0,0}\put(3001,-811){\vector( 1, 0){675}}
}%
\put(151,-886){\makebox(0,0)[lb]{\smash{{\SetFigFont{14}{16.8}{\rmdefault}{\mddefault}{\updefault}{\color[rgb]{0,0,0}$x_1$}%
}}}}
\put(151,-2836){\makebox(0,0)[lb]{\smash{{\SetFigFont{14}{16.8}{\rmdefault}{\mddefault}{\updefault}{\color[rgb]{0,0,0}$x_2$}%
}}}}
\put(1351,-2986){\makebox(0,0)[lb]{\smash{{\SetFigFont{14}{16.8}{\rmdefault}{\mddefault}{\updefault}{\color[rgb]{0,0,0}$h_{22}$}%
}}}}
\put(1351,-736){\makebox(0,0)[lb]{\smash{{\SetFigFont{14}{16.8}{\rmdefault}{\mddefault}{\updefault}{\color[rgb]{0,0,0}$h_{11}$}%
}}}}
\put(2626,-3586){\makebox(0,0)[lb]{\smash{{\SetFigFont{14}{16.8}{\rmdefault}{\mddefault}{\updefault}{\color[rgb]{0,0,0}$z_2$}%
}}}}
\put(2626, 14){\makebox(0,0)[lb]{\smash{{\SetFigFont{14}{16.8}{\rmdefault}{\mddefault}{\updefault}{\color[rgb]{0,0,0}$z_1$}%
}}}}
\put(2101,-1486){\makebox(0,0)[lb]{\smash{{\SetFigFont{14}{16.8}{\rmdefault}{\mddefault}{\updefault}{\color[rgb]{0,0,0}$h_{21}$}%
}}}}
\put(2101,-2236){\makebox(0,0)[lb]{\smash{{\SetFigFont{14}{16.8}{\rmdefault}{\mddefault}{\updefault}{\color[rgb]{0,0,0}$h_{12}$}%
}}}}
\put(3751,-886){\makebox(0,0)[lb]{\smash{{\SetFigFont{14}{16.8}{\rmdefault}{\mddefault}{\updefault}{\color[rgb]{0,0,0}$y_1$}%
}}}}
\put(3751,-2836){\makebox(0,0)[lb]{\smash{{\SetFigFont{14}{16.8}{\rmdefault}{\mddefault}{\updefault}{\color[rgb]{0,0,0}$y_2$}%
}}}}
\put(2878,-851){\makebox(0,0)[lb]{\smash{{\SetFigFont{12}{14.4}{\rmdefault}{\mddefault}{\updefault}{\color[rgb]{0,0,0}+}%
}}}}
{\color[rgb]{0,0,0}\put(2923,-786){\circle{150}}
}%
\end{picture}%

%% file: CompoundIC.latex
\setlength{\unitlength}{1184sp}%
\begingroup\makeatletter\ifx\SetFigFont\undefined%
\gdef\SetFigFont#1#2#3#4#5{%
  \reset@font\fontsize{#1}{#2pt}%
  \fontfamily{#3}\fontseries{#4}\fontshape{#5}%
  \selectfont}%
\fi\endgroup%
\begin{picture}(14113,10067)(4109,-33009)
\put(12662,-23302){\makebox(0,0)[b]{\smash{{\SetFigFont{9}{10.8}{\familydefault}{\mddefault}{\updefault}{\color[rgb]{0,0,0}$Y_{1\alpha}$}%
}}}}
{\color[rgb]{0,0,0}\thinlines
\put(11581,-26335){\circle*{134}}
}%
{\color[rgb]{0,0,0}\put(10863,-31282){\circle*{134}}
}%
{\color[rgb]{0,0,0}\put(11564,-29629){\circle*{134}}
}%
{\color[rgb]{0,0,0}\put(4809,-31282){\circle*{134}}
}%
{\color[rgb]{0,0,0}\put(4809,-24678){\circle*{134}}
}%
\thicklines
{\color[rgb]{0,0,0}\put(4489,-24672){\line( 1, 0){11338}}
\put(15827,-24672){\vector( 0, 1){708}}
}%
{\color[rgb]{0,0,0}\put(4489,-31286){\line( 1, 0){11338}}
\put(15827,-31286){\vector( 0,-1){709}}
}%
{\color[rgb]{0,0,0}\put(9922,-29633){\vector( 1, 0){2645}}
}%
{\color[rgb]{0,0,0}\put(15119,-29633){\line( 1, 0){1417}}
\put(16536,-29633){\vector( 0,-1){2362}}
}%
{\color[rgb]{0,0,0}\put(10867,-31286){\vector( 0,-1){709}}
}%
{\color[rgb]{0,0,0}\put(11575,-29633){\vector( 0,-1){2362}}
}%
{\color[rgb]{0,0,0}\put(9922,-26326){\vector( 1, 0){2645}}
}%
{\color[rgb]{0,0,0}\put(15119,-26326){\line( 1, 0){1417}}
\put(16536,-26326){\vector( 0, 1){2362}}
}%
{\color[rgb]{0,0,0}\put(10867,-24672){\vector( 0, 1){708}}
}%
{\color[rgb]{0,0,0}\put(11575,-26326){\vector( 0, 1){2362}}
}%
{\color[rgb]{0,0,0}\put(6035,-26301){\vector( 1, 0){1282}}
}%
{\color[rgb]{0,0,0}\put(4801,-31261){\line( 1, 4){1237.941}}
}%
{\color[rgb]{0,0,0}\put(10410,-23964){\framebox(1638,945){}}
}%
{\color[rgb]{0,0,0}\put(15371,-23964){\framebox(1637,945){}}
}%
{\color[rgb]{0,0,0}\put(12536,-27050){\framebox(2598,1417){}}
}%
{\color[rgb]{0,0,0}\put(7339,-27050){\framebox(2599,1417){}}
}%
{\color[rgb]{0,0,0}\put(7339,-30326){\framebox(2599,1417){}}
}%
{\color[rgb]{0,0,0}\put(12536,-30326){\framebox(2598,1417){}}
}%
{\color[rgb]{0,0,0}\put(15371,-32987){\framebox(1637,945){}}
}%
{\color[rgb]{0,0,0}\put(12028,-32515){\vector( 1, 0){1201}}
}%
{\color[rgb]{0,0,0}\put(10410,-32987){\framebox(1638,945){}}
}%
{\color[rgb]{0,0,0}\put(16988,-32515){\vector( 1, 0){1201}}
}%
{\color[rgb]{0,0,0}\put(16988,-23491){\vector( 1, 0){1201}}
}%
{\color[rgb]{0,0,0}\put(12028,-23491){\vector( 1, 0){1201}}
}%
{\color[rgb]{0,0,0}\put(4807,-24683){\line( 1,-4){1237.941}}
}%
{\color[rgb]{0,0,0}\put(6055,-29606){\vector( 1, 0){1282}}
}%
\put(4725,-31759){\makebox(0,0)[b]{\smash{{\SetFigFont{9}{10.8}{\familydefault}{\mddefault}{\updefault}{\color[rgb]{0,0,0}$X_2$}%
}}}}
\put(4725,-24483){\makebox(0,0)[b]{\smash{{\SetFigFont{9}{10.8}{\familydefault}{\mddefault}{\updefault}{\color[rgb]{0,0,0}$X_1$}%
}}}}
\put(11197,-23586){\makebox(0,0)[b]{\smash{{\SetFigFont{9}{10.8}{\familydefault}{\mddefault}{\updefault}{\color[rgb]{0,0,0}$f_{1\alpha}$}%
}}}}
\put(16158,-23586){\makebox(0,0)[b]{\smash{{\SetFigFont{9}{10.8}{\familydefault}{\mddefault}{\updefault}{\color[rgb]{0,0,0}$f_{1\beta}$}%
}}}}
\put(13812,-26483){\makebox(0,0)[b]{\smash{{\SetFigFont{9}{10.8}{\familydefault}{\mddefault}{\updefault}{\color[rgb]{0,0,0}$p_{S_{2\beta}|S_{2\alpha}}$}%
}}}}
\put(8615,-26483){\makebox(0,0)[b]{\smash{{\SetFigFont{9}{10.8}{\familydefault}{\mddefault}{\updefault}{\color[rgb]{0,0,0}$p_{S_{2\alpha}|X_2}$}%
}}}}
\put(8615,-29696){\makebox(0,0)[b]{\smash{{\SetFigFont{9}{10.8}{\familydefault}{\mddefault}{\updefault}{\color[rgb]{0,0,0}$p_{S_{1\alpha}|X_1}$}%
}}}}
\put(13812,-29696){\makebox(0,0)[b]{\smash{{\SetFigFont{9}{10.8}{\familydefault}{\mddefault}{\updefault}{\color[rgb]{0,0,0}$p_{S_{1\beta}|S_{1\alpha}}$}%
}}}}
\put(16158,-32609){\makebox(0,0)[b]{\smash{{\SetFigFont{9}{10.8}{\familydefault}{\mddefault}{\updefault}{\color[rgb]{0,0,0}$f_{2\beta}$}%
}}}}
\put(12662,-32326){\makebox(0,0)[b]{\smash{{\SetFigFont{9}{10.8}{\familydefault}{\mddefault}{\updefault}{\color[rgb]{0,0,0}$Y_{2\alpha}$}%
}}}}
\put(11197,-32609){\makebox(0,0)[b]{\smash{{\SetFigFont{9}{10.8}{\familydefault}{\mddefault}{\updefault}{\color[rgb]{0,0,0}$f_{2\alpha}$}%
}}}}
\put(17623,-32326){\makebox(0,0)[b]{\smash{{\SetFigFont{9}{10.8}{\familydefault}{\mddefault}{\updefault}{\color[rgb]{0,0,0}$Y_{2\beta}$}%
}}}}
\put(17623,-23302){\makebox(0,0)[b]{\smash{{\SetFigFont{9}{10.8}{\familydefault}{\mddefault}{\updefault}{\color[rgb]{0,0,0}$Y_{1\beta}$}%
}}}}
{\color[rgb]{0,0,0}\thinlines
\put(10863,-24680){\circle*{134}}
}%
\end{picture}%

%% file: nCompoundIC.latex
\setlength{\unitlength}{1184sp}%
\begingroup\makeatletter\ifx\SetFigFont\undefined%
\gdef\SetFigFont#1#2#3#4#5{%
  \reset@font\fontsize{#1}{#2pt}%
  \fontfamily{#3}\fontseries{#4}\fontshape{#5}%
  \selectfont}%
\fi\endgroup%
\begin{picture}(19454,10171)(4468,-33008)
\put(19726,-29761){\makebox(0,0)[b]{\smash{{\SetFigFont{9}{10.8}{\familydefault}{\mddefault}{\updefault}{\color[rgb]{0,0,0}$p_{S_{1\alpha_N}|S_{1\alpha_{(N-1)}}}$}%
}}}}
{\color[rgb]{0,0,0}\thinlines
\put(4809,-24678){\circle*{134}}
}%
\thicklines
{\color[rgb]{0,0,0}\put(6035,-26301){\vector( 1, 0){1282}}
}%
{\color[rgb]{0,0,0}\put(4801,-31261){\line( 1, 4){1237.941}}
}%
{\color[rgb]{0,0,0}\put(7339,-30326){\framebox(2599,1417){}}
}%
{\color[rgb]{0,0,0}\put(4807,-24683){\line( 1,-4){1237.941}}
}%
{\color[rgb]{0,0,0}\put(7339,-27050){\framebox(2599,1417){}}
}%
{\color[rgb]{0,0,0}\put(10651,-26326){\vector( 0, 1){2362}}
}%
{\color[rgb]{0,0,0}\put(9585,-23986){\framebox(1638,945){}}
}%
{\color[rgb]{0,0,0}\put(11411,-27053){\framebox(2598,1417){}}
}%
{\color[rgb]{0,0,0}\put(10126,-24747){\vector( 0, 1){708}}
}%
{\color[rgb]{0,0,0}\put(6055,-29606){\vector( 1, 0){1282}}
}%
{\color[rgb]{0,0,0}\put(13646,-23986){\framebox(1637,945){}}
}%
{\color[rgb]{0,0,0}\put(15338,-23536){\vector( 1, 0){1201}}
}%
{\color[rgb]{0,0,0}\put(4501,-24661){\line( 1, 0){9750}}
\put(14251,-24661){\vector( 0, 1){675}}
}%
{\color[rgb]{0,0,0}\put(14026,-26386){\line( 1, 0){600}}
\put(14626,-26386){\vector( 0, 1){2400}}
}%
{\color[rgb]{0,0,0}\put(9976,-26386){\vector( 1, 0){1425}}
}%
{\color[rgb]{0,0,0}\put(9585,-32986){\framebox(1638,945){}}
}%
{\color[rgb]{0,0,0}\put(11203,-32536){\vector( 1, 0){1201}}
}%
{\color[rgb]{0,0,0}\put(10126,-31361){\vector( 0,-1){709}}
}%
{\color[rgb]{0,0,0}\put(13646,-32881){\framebox(1637,945){}}
}%
{\color[rgb]{0,0,0}\put(15300,-32536){\vector( 1, 0){1201}}
}%
{\color[rgb]{0,0,0}\put(9976,-29611){\line( 1, 0){1425}}
}%
{\color[rgb]{0,0,0}\put(10651,-29611){\vector( 0,-1){2400}}
}%
{\color[rgb]{0,0,0}\put(4501,-31336){\line( 1, 0){9750}}
\put(14251,-31336){\vector( 0,-1){600}}
}%
{\color[rgb]{0,0,0}\put(21301,-29611){\line( 1, 0){600}}
\put(21901,-29611){\vector( 0,-1){2325}}
}%
{\color[rgb]{0,0,0}\put(21301,-26311){\line( 1, 0){600}}
\put(21901,-26311){\vector( 0, 1){2400}}
}%
{\color[rgb]{0,0,0}\put(21001,-32806){\framebox(1637,945){}}
}%
{\color[rgb]{0,0,0}\put(22650,-32461){\vector( 1, 0){1201}}
}%
{\color[rgb]{0,0,0}\put(21001,-23956){\framebox(1637,945){}}
}%
{\color[rgb]{0,0,0}\put(22688,-23611){\vector( 1, 0){1201}}
}%
{\color[rgb]{0,0,0}\put(18376,-24661){\line( 1, 0){3075}}
\put(21451,-24661){\vector( 0, 1){675}}
}%
{\color[rgb]{0,0,0}\put(18301,-31336){\line( 1, 0){3225}}
\put(21526,-31336){\vector( 0,-1){525}}
}%
{\color[rgb]{0,0,0}\put(11203,-23536){\vector( 1, 0){1201}}
}%
{\color[rgb]{0,0,0}\multiput(14251,-24661)(750.00000,0.00000){6}{\line( 1, 0){375.000}}
}%
{\color[rgb]{0,0,0}\put(18001,-27061){\framebox(3300,1425){}}
}%
{\color[rgb]{0,0,0}\put(18001,-30361){\framebox(3300,1425){}}
}%
{\color[rgb]{0,0,0}\put(17476,-29611){\line( 1, 0){525}}
}%
{\color[rgb]{0,0,0}\multiput(14251,-31336)(736.36364,0.00000){6}{\line( 1, 0){368.182}}
}%
{\color[rgb]{0,0,0}\multiput(14626,-26386)(857.14286,0.00000){4}{\line( 1, 0){428.571}}
}%
{\color[rgb]{0,0,0}\put(17626,-26386){\line( 1, 0){375}}
}%
{\color[rgb]{0,0,0}\put(14026,-29611){\line( 1, 0){600}}
\put(14626,-29611){\vector( 0,-1){2325}}
}%
{\color[rgb]{0,0,0}\multiput(14626,-29611)(814.28571,0.00000){4}{\line( 1, 0){407.143}}
}%
{\color[rgb]{0,0,0}\put(11411,-30361){\framebox(2598,1417){}}
}%
\put(4725,-31759){\makebox(0,0)[b]{\smash{{\SetFigFont{9}{10.8}{\familydefault}{\mddefault}{\updefault}{\color[rgb]{0,0,0}$X_2$}%
}}}}
\put(4725,-24483){\makebox(0,0)[b]{\smash{{\SetFigFont{9}{10.8}{\familydefault}{\mddefault}{\updefault}{\color[rgb]{0,0,0}$X_1$}%
}}}}
\put(8615,-26483){\makebox(0,0)[b]{\smash{{\SetFigFont{9}{10.8}{\familydefault}{\mddefault}{\updefault}{\color[rgb]{0,0,0}$p_{S_{2\alpha_1}|X_2}$}%
}}}}
\put(8615,-29696){\makebox(0,0)[b]{\smash{{\SetFigFont{9}{10.8}{\familydefault}{\mddefault}{\updefault}{\color[rgb]{0,0,0}$p_{S_{1\alpha_1}|X_1}$}%
}}}}
\put(10426,-23686){\makebox(0,0)[b]{\smash{{\SetFigFont{9}{10.8}{\familydefault}{\mddefault}{\updefault}{\color[rgb]{0,0,0}$f_{1\alpha_1}$}%
}}}}
\put(14476,-23686){\makebox(0,0)[b]{\smash{{\SetFigFont{9}{10.8}{\familydefault}{\mddefault}{\updefault}{\color[rgb]{0,0,0}$f_{1\alpha_2}$}%
}}}}
\put(10351,-32611){\makebox(0,0)[b]{\smash{{\SetFigFont{9}{10.8}{\familydefault}{\mddefault}{\updefault}{\color[rgb]{0,0,0}$f_{2\alpha_1}$}%
}}}}
\put(14401,-32611){\makebox(0,0)[b]{\smash{{\SetFigFont{9}{10.8}{\familydefault}{\mddefault}{\updefault}{\color[rgb]{0,0,0}$f_{2\alpha_2}$}%
}}}}
\put(21826,-32536){\makebox(0,0)[b]{\smash{{\SetFigFont{9}{10.8}{\familydefault}{\mddefault}{\updefault}{\color[rgb]{0,0,0}$f_{2\alpha_N}$}%
}}}}
\put(12676,-26536){\makebox(0,0)[b]{\smash{{\SetFigFont{9}{10.8}{\familydefault}{\mddefault}{\updefault}{\color[rgb]{0,0,0}$p_{S_{2\alpha_2}|S_{2\alpha_1}}$}%
}}}}
\put(12676,-29761){\makebox(0,0)[b]{\smash{{\SetFigFont{9}{10.8}{\familydefault}{\mddefault}{\updefault}{\color[rgb]{0,0,0}$p_{S_{1\alpha_2}|S_{1\alpha_1}}$}%
}}}}
\put(11926,-23236){\makebox(0,0)[b]{\smash{{\SetFigFont{9}{10.8}{\familydefault}{\mddefault}{\updefault}{\color[rgb]{0,0,0}$Y_{1\alpha_1}$}%
}}}}
\put(15976,-23236){\makebox(0,0)[b]{\smash{{\SetFigFont{9}{10.8}{\familydefault}{\mddefault}{\updefault}{\color[rgb]{0,0,0}$Y_{1\alpha_2}$}%
}}}}
\put(21826,-23686){\makebox(0,0)[b]{\smash{{\SetFigFont{9}{10.8}{\familydefault}{\mddefault}{\updefault}{\color[rgb]{0,0,0}$f_{1\alpha_N}$}%
}}}}
\put(23326,-23236){\makebox(0,0)[b]{\smash{{\SetFigFont{9}{10.8}{\familydefault}{\mddefault}{\updefault}{\color[rgb]{0,0,0}$Y_{1\alpha_N}$}%
}}}}
\put(11926,-32161){\makebox(0,0)[b]{\smash{{\SetFigFont{9}{10.8}{\familydefault}{\mddefault}{\updefault}{\color[rgb]{0,0,0}$Y_{2\alpha_1}$}%
}}}}
\put(15976,-32161){\makebox(0,0)[b]{\smash{{\SetFigFont{9}{10.8}{\familydefault}{\mddefault}{\updefault}{\color[rgb]{0,0,0}$Y_{2\alpha_2}$}%
}}}}
\put(23326,-32011){\makebox(0,0)[b]{\smash{{\SetFigFont{9}{10.8}{\familydefault}{\mddefault}{\updefault}{\color[rgb]{0,0,0}$Y_{2\alpha_N}$}%
}}}}
\put(19726,-26536){\makebox(0,0)[b]{\smash{{\SetFigFont{9}{10.8}{\familydefault}{\mddefault}{\updefault}{\color[rgb]{0,0,0}$p_{S_{2\alpha_N}|S_{2\alpha_{(N-1)}}}$}%
}}}}
{\color[rgb]{0,0,0}\thinlines
\put(4809,-31282){\circle*{134}}
}%
\end{picture}%